\documentclass[12pt]{article}

\pdfoutput=1
\usepackage{color}
\usepackage{epsfig, palatino}
\usepackage{pstricks,pst-node,pst-tree}
\usepackage{epic}
\usepackage{mathrsfs}
\usepackage{ae} 
\usepackage[T1]{fontenc}
\usepackage[ansinew]{inputenc}
\usepackage{amsmath}
\usepackage{amssymb}
\usepackage{comment}
\usepackage{graphicx}
\usepackage{ulem}
\usepackage{xcolor}
\definecolor{darkblue}{cmyk}{0.9,0.9,0,0}
\definecolor{c1}{RGB}{236, 0, 11}
\definecolor{c2}{RGB}{0, 163, 244}
\definecolor{c3}{RGB}{170, 4, 212}
\definecolor{c4}{RGB}{255, 153, 0}
\definecolor{c5}{RGB}{45, 224, 0}
\usepackage[colorlinks=true,linkcolor=darkblue,citecolor=darkblue,urlcolor=darkblue]{hyperref}
\usepackage{cite}
\usepackage{hyperref}
\usepackage{wasysym}
\usepackage{varioref}
\usepackage{makeidx}
\usepackage[english]{babel}
\usepackage{simplewick}
\usepackage{array}
\usepackage{multirow}

\usepackage[font={small}]{caption}

\newcommand{\beq}{\begin{equation}}
\newcommand{\eeq}{\end{equation}}
\newcommand{\beqq}{\begin{equation*}}
\newcommand{\eeqq}{\end{equation*}}
\newcommand\beqa{\begin{eqnarray}}
\newcommand\eeqa{\end{eqnarray}}
\newcommand\beqaa{\begin{eqnarray*}}
\newcommand\eeqaa{\end{eqnarray*}}
\newcommand\bea{\begin{array}}
\newcommand\eea{\end{array}}

\newcommand{\neqa}{\nonumber\end{eqnarray}}

\renewcommand{\d}{\partial}

\newcommand{\<}{{\langle}}
\renewcommand{\>}{{\rangle}}

\newcommand{\re}{\relax{\rm I\kern-.18em R}}

\renewcommand{\sp}{p\hspace{-.40em}/}

\definecolor{darkgreen}{rgb}{0.0, 0.45, 0.0}

\def\XXint#1#2#3{{\setbox0=\hbox{$#1{#2#3}{\int}$}
\vcenter{\hbox{$#2#3$}}\kern-.5\wd0}}

\def\su2{{SU(2)}}

\def\[{\left[}
\def\]{\right]}

\def\({\left(}
\def\){\right)}
\def\[{\left[}
\def\]{\right]}

\def\<{\langle}
\def\>{\rangle}

\def\i2{\frac{i}{2}}

\def\spi{\relax{\rm \pi\kern-0.5em /}}
\def\sA{\relax{\rm A\kern-0.5em /}}
\def\sp{\relax{\rm p\kern-0.5em /}}
\def\sd{\relax{\rm \d\kern-0.5em /}}
\def\sk{\relax{\rm k\kern-0.5em /}}
\def\sn{\relax{\rm n\kern-0.5em /}}
\def\sl{\relax{\rm l\kern-0.5em /}}
\def\sP{\relax{\rm P\kern-0.7em /}}
\def\sBethe{\relax{\rm \Bethe\kern-0.5em /}}

\def\2F1{\,_2{\rm F}_1}

\makeindex

\newcommand\blfootnote[1]{%
	\begingroup
	\renewcommand\thefootnote{}\footnote{\hspace{-6mm}#1}%
	\addtocounter{footnote}{-1}%
	\endgroup
}

\usepackage{tikz}
\usetikzlibrary{arrows}
\usetikzlibrary{arrows.meta}
\usetikzlibrary{decorations.pathmorphing}
\usetikzlibrary{snakes}
\usetikzlibrary{decorations.pathreplacing,decorations.markings,decorations.pathmorphing}\usetikzlibrary{snakes}
\usetikzlibrary{decorations.pathreplacing,decorations.markings,decorations.pathmorphing}
\usetikzlibrary{calc}
	
\begin{document}

\thispagestyle{empty}

\renewcommand{\thefootnote}{\fnsymbol{footnote}}
\setcounter{page}{1}
\setcounter{footnote}{0}
\setcounter{figure}{0}


\begin{center}
	$$$$
	{\Large\textbf{\mathversion{bold}
			Two-loop integrals of half-BPS six-point functions on a line}}
	
	\vspace{1.0cm}

	\textrm{Ricardo Rodrigues  }
	\blfootnote{\tt{up201805119@edu.fc.up.pt}}
	\\ \vspace{1.2cm}
	\footnotesize{\textit{
			Centro de F$\acute{\imath}$sica do Porto, Departamento de F$\acute{\imath}$sica e Astronomia,
			Faculdade de Ci$\hat{e}$ncias da Universidade do Porto, Rua do Campo Alegre 687, 4169-007 Porto, Portugal\\
		}  
		\vspace{4mm}
	}

	\par\vspace{1.5cm}
	
	\textbf{Abstract}\vspace{2mm}
\end{center}
We evaluated all two-loop conformal integrals of scalar half-BPS six-point functions in $\mathcal{N} = 4$ SYM restricted to a configuration where all points lie on a line. Moreover, we also computed some of these integrals in the kinematical limit where adjacent points become null-separated. Our results can serve as cross-checks for future works which obtain these integrals for general kinematics or by different methods such as integrability.

\noindent

\setcounter{page}{1}
\renewcommand{\thefootnote}{\arabic{footnote}}
\setcounter{footnote}{0}

\setcounter{tocdepth}{2}

\def\nref#1{{(\ref{#1})}}

\newpage

\tableofcontents

\parskip 5pt plus 1pt   \jot = 1.5ex

\section{Introduction} \label{sec:Introduction}
The applicability of conformal field theories ranges from the study of critical phenomena to that of quantum gravity. As such, these theories have been intensively studied over the past few decades. Among these, the $\mathcal{N} = 4$ supersymmetric Yang-Mills theory plays a special role not only because it is integrable in the planar limit but also because it is one of the few known examples of an integrable theory in four dimensions. As is the case with most CFT, correlation functions of local operators are one of the most interesting observables to study. For this theory in particular it has been shown to exist certain dualities between some of these correlation functions, scattering amplitudes and Wilson loops \cite{Eden:2010zz,Alday:2010zy,Eden:2010ce,Eden:2011yp}. 

The $\mathcal{N} = 4$ SYM theory has also appeared in the first example of the AdS/CFT duality, having been shown to be dual to a type IIB string theory in a $\text{AdS}_5 \times \text{S}_5$ geometry\cite{Maldacena:1997re}. Since then, this theory has proved to be of particular usefulness in having a better understanding of strong coupling phenomena, given that AdS/CFT is a strong/weak coupling duality. Within its spectrum a special kind of operators called half-BPS operators are dual to Kaluza-Klein spherical harmonics of the graviton on the sphere. As such, the analysis of correlators of such operators can offer us important information about the internal sector of the dual bulk theory and how its information is encoded in the CFT side of the duality. 

The above are some of the reasons why we are concerned with studying correlation functions in $\mathcal{N} = 4$ SYM, in particular of half-BPS operators. These, however, are not always easy to obtain. In fact, although some four-point functions of this theory have been studied both for weak \cite{Eden:2000mv,Eden:2011we,Chicherin:2015edu,Georgoudis:2017meq} and strong \cite{DHoker:1999kzh,Arutyunov:2000py,Goncalves:2014ffa,Rastelli:2016nze,Alday:2017xua,Binder:2019jwn} coupling, where the computations have been taken to a relatively high number of loops, the situation for higher-point functions is not as promising. Five and six-point functions, however, are somewhat of an exception to this as they have been analyzed relatively recently. In particular, the integrands of five-point functions are known up to three loops \cite{Bargheer:2022sfd}, while the integrals have been  explicitly obtained up to two loops when null limits are taken or when we restrict the points to a plane \cite{Bercini:2024pya}. There have also been some bootstrap approaches to certain five-point functions of this theory at strong coupling in \cite{Goncalves:2019znr,Goncalves:2023oyx}. On the other hand, the integrands of six-point functions have been obtained up to two loops \cite{Eden:2010zz,Eden:2010ce,Bargheer:2022sfd} but we still have not been able to obtain the integrated contributions to the corrections.

Here our main concern will be the two loop conformal integrals associated to six-point functions of scalar single-trace half-BPS operators. In particular, since the complexity of these integrals for general kinematics is high enough that we cannot evaluate them explicitly, we will restrict the configuration to a line geometry which simplifies things. Additionally, we will also evaluate some of these integrals when null limits are taken.

The structure of this paper is as follows. In section \ref{sec:Correlation functions} we introduce the single-trace half-BPS operators, giving special focus to twenty-prime operators, their correlation functions and respective properties. Section \ref{sec:Integrals} describes the procedure to compute the integrals on the line geometry and presents the different types of two loop six-point conformal integrals. Additionally, it tackles some important points about the line geometry and the null limit cases. In section \ref{sec:Results and analysis} we comment the results regarding the line configuration and then move on to present the explicit outcome of evaluating one of the integrals in the light-cone limit. In section \ref{sec:Discussion} we make a general assessment of our work and elaborate on possible future directions. Appendix \ref{sec:Graph Polynomials} contains explicit expressions of quantities concerning the integrand of a conformal integral considered in section \ref{subsec:Line geometry}. In appendix \ref{sec:Asymptotic expanded integrals} we briefly explain how to generalize a result for asymptotic expanded integrals which was necessary to evaluate the integral considered in section \ref{sec:Results and analysis}.

\section{Correlation functions} \label{sec:Correlation functions}
Here, we consider the four dimensional $\mathcal{N} = 4$ SYM theory, with gauge group $SU(N_c)$, where $N_c$ is the number of colors, which among other fields contains six bosonic scalars denoted by $\Phi^{I}=(\phi^1,\dots,\phi^6)$. Moreover, this theory has an associated coupling constant $g_{\text{YM}}$ which is related with $N_c$ by the so-called 't Hooft coupling
\begin{equation}
	\lambda = g_{\text{YM}}^2 \, N_c \,\, .
	\label{eq:t Hooft coupling def}
\end{equation}
Within the spectrum of this theory there is a class of important operators called the scalar single-trace half-BPS operators, which transform in the symmetric and traceless representation of the $SO(6)_R \simeq SU(4)_R$ global symmetry group. They are generally written as
\begin{equation}
	\mathcal{O}_k(x,Y) = Y^{i_1} \dots Y^{i_k} \, \text{tr}\left(\phi^{i_1}(x) \dots \phi^{i_k}(x)\right) \,\, ,
	\label{eq:single-trace half-BPS ops def}
\end{equation}
where $k \ge 2$ is an integer, $Y^i$ are six-dimensional null polarization vectors ($Y^i \cdot Y^i = 0$) contracting with the $SO(6)_R$ indices, and the trace is taken over the indices of $SU(N_c)$, which are omitted for simplicity. Such operators $\mathcal{O}_k$ have conformal dimension $\Delta_{\mathcal{O}_k} = k$.

An interesting property of these operators is that their two- and three-point functions are protected from quantum corrections by supersymmetry \cite{Intriligator:1998ig}, which means they remain unchanged independently of the theory's coupling. On the other hand, while their higher-point functions no longer enjoy this property they are of extreme importance to us as they encode useful information about the theory itself and its holographic dual.

In general, any planar $n$-point function of scalar single-trace half-BPS operators is given by
\begin{equation}
	\langle \mathcal{O}_{k_1} \, \mathcal{O}_{k_2} \dots \mathcal{O}_{k_n} \rangle = \sum_{\ell = 0}^{\infty} \lambda^{2 \ell} \, G_n^{(\ell)} \,,
	\label{eq:perturbative expansion of n-pt function of half-BPS}
\end{equation}
while the non-planar contributions are subleading in the large $N_c$ limit. In the above expression $G_n^{(0)}$ concerns the free correlator, whereas the terms with $\ell \ge 1$ account for loop corrections, which are given by specific spacetime integrals. In practice, the integrands of these corrections can be computed through the Lagrangian insertion method \cite{Intriligator:1998ig,Eden:2010ce,Eden:2010zz,Caron-Huot:2010ryg,Eden:2011we}, in which they are given by $(n+\ell)$-point functions with $\ell$ insertions of the Lagrangian density operator. However, as we start considering more points and/or higher loops computing the integrands becomes harder as it requires the evaluation of higher-point correlators. Additionally, once we have the integrands we still have to compute the integrals which can be quite challenging.

Inside this special class of operators one of the most important and studied ones is the lowest length half-BPS operator ($k=2$), which is usually denoted by $20'$ operator
\begin{equation}
	O_2(x,Y) \equiv O_{20'}(x,Y) = Y^{i_1} \, Y^{i_2} \, \text{tr}\left(\phi^{i_1}(x) \, \phi^{i_2}(x)\right) \,\, .
	\label{eq:20' op def}
\end{equation}
It belongs to the stress-tensor supermultiplet along with the stress-tensor itself, the Lagrangian density operator and others. Given the above definition, we expect the correlation functions of these operators to be given in terms of polynomials of polarization vectors that have weight two in each $Y_i$.

The two- and three-point functions of the twenty-prime operators are completely fixed by conformal symmetry and supersymmetry to be given by
\begin{align}
	&\langle \mathcal{O}_{20'}(x_1,Y_1) \, \mathcal{O}_{20'}(x_2,Y_2) \rangle = \left(\frac{Y_{12}}{x_{12}^2}\right)^2 \, ,	\label{eq:20' two pt function} \\
	&\langle \mathcal{O}_{20'}(x_1,Y_1) \, \mathcal{O}_{20'}(x_2,Y_2) \, \mathcal{O}_{20'}(x_3,Y_3) \rangle = \frac{C_{20' 20' 20'} \, Y_{12} Y_{13} Y_{23} }{(x_{12}^2 x_{13}^2 x_{23}^2)} \,\, ,
	\label{eq:20' three pt function}
\end{align} 
where we have used the notation $x_{ij}^2 \equiv (x_i - x_j)^2$, $Y_{ij} \equiv Y_i \cdot Y_j$, and where $C_{20' 20' 20'}$ is the structure constant of its three-point function, which is independent of the theory's coupling. On the other hand, the higher-point functions, which are not protected by supersymmetry, are completely non-trivial. 

Let us then start by considering the four-point function, which is the lowest of the higher-point correlators. From a simple analysis it is possible to infer that the polynomial in the polarization vectors can contain up to two different structures. In particular, all the structures we can find are

\begin{figure}[h!]
	\centering
	\begin{tikzpicture}
			\node[circle, fill, inner sep=2pt, label={[xshift=-4pt]left:1}] at (-4,1) {};
			\node[circle, fill, inner sep=2pt,label={[xshift=-4pt]left:2}] at (-4,-1) {};
			\node[circle, fill, inner sep=2pt,label={[xshift=4pt]right:4}] at (-2,1) {};
			\node[circle, fill, inner sep=2pt,label={[xshift=4pt]right:3}] at (-2,-1) {};
			
			\draw[black,thin] (-4.05,1)--(-4.05,-1);
			\draw[black,thin] (-4.05,1)--(-4.05,-1);
			\draw[black,thin] (-3.95,1)--(-3.95,-1);
			\draw[black,thin] (-3.95,1)--(-3.95,-1);
			\draw[black,thin] (-2.05,1)--(-2.05,-1);
			\draw[black,thin] (-2.05,1)--(-2.05,-1);
			\draw[black,thin] (-1.95,1)--(-1.95,-1);
			\draw[black,thin] (-1.95,1)--(-1.95,-1);
			
			\node[circle, fill, inner sep=2pt, label={[xshift=4pt]right:4}] at (4,1) {};
			\node[circle, fill, inner sep=2pt,label={[xshift=4pt]right:3}] at (4,-1) {};
			\node[circle, fill, inner sep=2pt,label={[xshift=-4pt]left:1}] at (2,1) {};
			\node[circle, fill, inner sep=2pt,label={[xshift=-4pt]left:2}] at (2,-1) {};
			
			\draw[black,thin] (4,1)--(4,-1);
			\draw[black,thin] (4,-1)--(2,-1);
			\draw[black,thin] (2,-1)--(2,1);
			\draw[black,thin] (2,1)--(4,1);
	\end{tikzpicture}
	\caption{Diagrams representing the $Y_{ij}$ structures of the twenty-prime four-point function.}
	\label{fig:diagram structures four-point function}
\end{figure}
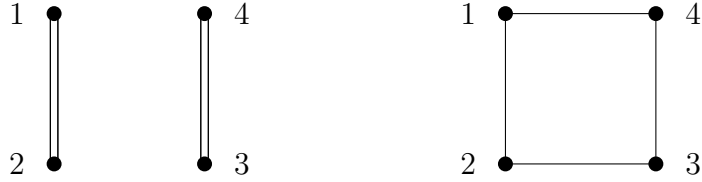

\begin{equation}
	Y_{12}^2 \, Y_{34}^2 \,\, , \hspace{0.5cm} Y_{12} \, Y_{23} \, Y_{34} \, Y_{41} \,\, ,
	\label{eq:possible structures four pt function}
\end{equation}
which correspond to the diagrams in figure \ref{fig:diagram structures four-point function}, plus the ones obtained from permuting the positions. As such, we can write the four-point function as
\begin{equation}
	\langle \mathcal{O}_{20'}(x_1,Y_1) \, \mathcal{O}_{20'}(x_2,Y_2) \, \mathcal{O}_{20'}(x_3,Y_3) \, \mathcal{O}_{20'}(x_4,Y_4) \rangle = \left(\frac{Y_{12} \, Y_{34}}{x_{12}^2 \, x_{34}^2}\right)^2 \, \mathcal{G}_4(u,v;\sigma,\tau) \, ,
	\label{eq:20' four point function}
\end{equation}
where the conformal and the R-symmetry\footnote{R-symmetry is a symmetry that is present in supersymmetric theories and which is responsible by transforming the different supercharges into each other. In this case, it will be equivalent to the SO(6) symmetry associated with rotations among the six scalar fields $\Phi^I$.} cross-ratios are respectively defined as
\begin{equation}
	\begin{aligned}
		&u = \frac{x_{12}^2 x_{34}^2}{x_{13}^2 x_{24}^2} = z \bar{z} \,, \hspace{0.3cm} v=\frac{x_{14}^2 x_{23}^2}{x_{13}^2 x_{24}^2} = (1-z)(1-\bar{z}) \, , \\
		&\sigma = \frac{Y_{12} Y_{34}}{Y_{13} Y_{24}} = \alpha \bar{\alpha} \,, \hspace{0.3cm} \tau=\frac{Y_{14} Y_{23}}{Y_{13} Y_{24}} = (1-\alpha)(1-\bar{\alpha}) \, .
	\end{aligned}
	\label{eq:four point conformal and R cross-ratios}
\end{equation}
More can be said about the cross-ratios function $\mathcal{G}$. Indeed, superconformal Ward identities impose extra non-trivial conditions on it \cite{Eden:2000bk,Nirschl:2004pa,Rastelli:2016nze} which in due turn enforce it to have a particular structure, namely
\begin{equation}
	\mathcal{G}_4(z,\bar{z};\alpha,\bar{\alpha}) = \mathcal{G}_4^{(0)}(z,\bar{z};\alpha,\bar{\alpha}) + R \, \mathcal{F}_4(z,\bar{z}) \, ,
	\label{eq:structure of 20' four pt cross-ratios function}
\end{equation}
where $\mathcal{G}_4^{(0)}(z,\bar{z};\alpha,\bar{\alpha})$ is the free part of the correlator $G_4^{(0)}$ with the position-dependent prefactor extracted, $R$ is given by
\begin{equation}
	R = \frac{(1 - z \, \alpha) \, (1 - \bar{z} \, \alpha) \, (1 - z \, \bar{\alpha}) \, (1 - \bar{z} \, \bar{\alpha})}{(1-z) \, (1-\bar{z}) \, \alpha^2 \, \bar{\alpha}^2} \, ,
	\label{eq:R factor of 20' four pt function}
\end{equation}
and $\mathcal{F}_4(z,\bar{z})$ contains all the loop-corrections mentioned in (\ref{eq:perturbative expansion of n-pt function of half-BPS}) also without the prefactor. This latter function is important in the sense that it is contains all the dynamical information.

When it comes to the four-point function of twenty-prime operators, this comprises one of the most well studied four-point functions in all conformal field theories. Indeed, the integrands of the corrections for this object are known in the weak coupling perturbative regime up to ten loops \cite{Eden:2012tu,Bourjaily:2016evz}. In spite of this, the conformal integrals have only been evaluated up to three loops for fixed cross-ratios \cite{Drummond:2013nda} and up to five loops in the Euclidean OPE limit \cite{Georgoudis:2017meq}. As for the strong coupling regime, the first few orders of the $1/N$ corrections have been obtained by means of conformal bootstrap methods \cite{Alday:2017vkk,Alday:2017xua,Alday:2018pdi,Caron-Huot:2018kta}.

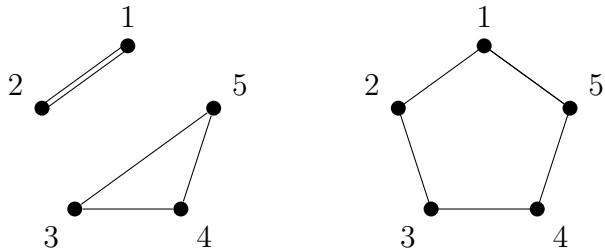
\begin{figure}[t!]
	\centering
	\begin{tikzpicture}[scale=1.2]
		\foreach \angle/\label/\num in {90/1/1, 162/2/2, 234/3/3, 306/4/4, 18/5/5}
		{
			\node[fill, circle, inner sep=2pt, label={\angle:\num}] (\num) at (\angle:1) {};
		}
		\draw[thin] ($(1) + (0.05,0)$) -- ($(2) + (0.05,0)$);
		\draw[thin] ($(1) - (0.05,0)$) -- ($(2) - (0.05,0)$);
		\draw[thin] (3) -- (4);
		\draw[thin] (4) -- (5);
		\draw[thin] (3) -- (5);
	\end{tikzpicture}
	\hspace{1cm} 
	\begin{tikzpicture}[scale=1.2]
		\foreach \angle/\label/\num in {90/1/1, 162/2/2, 234/3/3, 306/4/4, 18/5/5}
		{
			\node[fill, circle, inner sep=2pt, label={\angle:\num}] (\num) at (\angle:1) {};
			\draw[thin] (\angle:1) -- ({\angle+72}:1); 
		}
		\draw[thin] (1) -- (5);
	\end{tikzpicture}
	\caption{Diagrams representing the $Y_{ij}$ structures of the twenty-prime five-point function.}
	\label{fig:diagram structures five-point function}
\end{figure}

The five-point function of these same operators is also interesting to look at. Performing a similar analysis as for the four-point case we can see that the allowed structures in the null polarization vectors are
\begin{equation}
	Y_{12}^2 Y_{34} Y_{45} Y_{53} \,\, , \,\,\,\, Y_{12} Y_{23} Y_{34} Y_{45} Y_{51}  \, , 
	\label{eq:possible structures five pt function}
\end{equation}
corresponding to the diagrams of figure \ref{fig:diagram structures five-point function},
together with the ones obtained from indices permutations. Consequently, we can write this correlation function as
\begin{equation}
	\langle \mathcal{O}_{20'}(x_1,Y_1) \dots \mathcal{O}_{20'}(x_5,Y_5) \rangle = \frac{x_{13}^2 Y_{12}^2 Y_{34}^2 Y_{15} Y_{35}}{x_{12}^4 x_{34}^4 x_{15}^2 x_{35}^2 Y_{13}} \, \mathcal{G}_5(u_i;\sigma_i) \, ,
	\label{eq:20' five point function}
\end{equation}
which depends on five conformal cross-ratios $u_1, \dots , u_5$ defined as
\begin{equation}
	u_1 = \frac{x_{12}^2 x_{35}^2}{x_{13}^2 x_{25}^2} \,\, , \hspace{0.5cm} u_{i+1} = \left.u_i\right\vert_{x_j \rightarrow x_{j+1}} \, ,
	\label{eq:five point cross-ratios}
\end{equation}
where the index $j$ is taken modulo $5$, and on the R-symmetry cross-ratios $\sigma_1, \dots, \sigma_5$ given as
\begin{equation}
	\sigma_1 = \frac{Y_{12} Y_{35}}{Y_{13} Y_{25}} \,\, , \hspace{0.5cm} \sigma_{i+1} = \left.\sigma_i\right\vert_{Y_j \rightarrow Y_{j+1}} \, ,
	\label{eq:five point R-symmetry cross-ratios}
\end{equation}
with $j$ also taken modulo $5$. Similarly, the cross-ratios function $\mathcal{G}_5$ also needs to satisfy superconformal Ward identities like in the four-point case. However, as far as we know no explicit solution was found for five points and we do not know if it should have a similar structure to (\ref{eq:structure of 20' four pt cross-ratios function}) or not.

Regarding the status of the five-point function corrections, the current situation is as referred previously. In particular, while the integrands have been obtained up to three loops \cite{Bargheer:2022sfd}, the integrals have only been computed up to $2$ loops for special cases, such as when null limits are taken or when all the points are set to the plane \cite{Bercini:2024pya}.

\begin{figure}[h!]
	\centering	
		\begin{tikzpicture}[scale=1.2]
		\foreach \angle/\label/\num in {90/1/1, 150/2/2, 210/3/3, 270/4/4, 330/5/5, 30/6/6}
		{
			\node[fill, circle, inner sep=2pt, label={\angle:\num}] (\num) at (\angle:1) {};
		}
		\draw[thin] ($(1) + (0.05,0)$) -- ($(2) + (0.05,0)$);
		\draw[thin] ($(1) - (0.05,0)$) -- ($(2) - (0.05,0)$);
		
		\draw[thin] ($(3) + (0.05,0)$) -- ($(4) + (0.05,0)$);
		\draw[thin] ($(3) - (0.05,0)$) -- ($(4) - (0.05,0)$);
		
		\draw[thin] ($(5) + (0.03,0)$) -- ($(6) + (0.03,0)$);
		\draw[thin] ($(5) - (0.03,0)$) -- ($(6) - (0.03,0)$);
	\end{tikzpicture}
	\hspace{0.5cm} 
	\begin{tikzpicture}[scale=1.2]
		\foreach \angle/\label/\num in {90/1/1, 150/2/2, 210/3/3, 270/4/4, 330/5/5, 30/6/6}
		{
			\node[fill, circle, inner sep=2pt, label={\angle:\num}] (\num) at (\angle:1) {};
		}
		\draw[thin] ($(1) + (0.05,0)$) -- ($(2) + (0.05,0)$);
		\draw[thin] ($(1) - (0.05,0)$) -- ($(2) - (0.05,0)$);
		
		\foreach \x/\y in {3/4, 4/5, 5/6, 6/3}
		{
			\draw[thin] (\x) -- (\y);
		}
	\end{tikzpicture}
	\hspace{0.5cm} 
	\begin{tikzpicture}[scale=1.2]
		\foreach \angle/\label/\num in {90/1/1, 150/2/2, 210/3/3, 270/4/4, 330/5/5, 30/6/6}
		{
			\node[fill, circle, inner sep=2pt, label={\angle:\num}] (\num) at (\angle:1) {};
			\draw[thin] (\angle:1) -- ({\angle+60}:1); 
		}
		\draw[thin] (1) -- (6);
	\end{tikzpicture}	
	\hspace{0.5cm} 
	\begin{tikzpicture}[scale=1.2]
		\foreach \angle/\label/\num in {90/1/1, 150/2/2, 210/3/3, 270/4/4, 330/5/5, 30/6/6}
		{
			\node[fill, circle, inner sep=2pt, label={\angle:\num}] (\num) at (\angle:1) {};
		}
		\draw[thin] (1) -- (2);
		\draw[thin] (2) -- (3);
		\draw[thin] (3) -- (1);
		
		\draw[thin] (4) -- (5);
		\draw[thin] (5) -- (6);
		\draw[thin] (6) -- (4);
	\end{tikzpicture}
	\caption{Diagrams representing the $Y_{ij}$ structures of the twenty-prime six-point function.}
	\label{fig:diagram structures six-point function}
\end{figure}
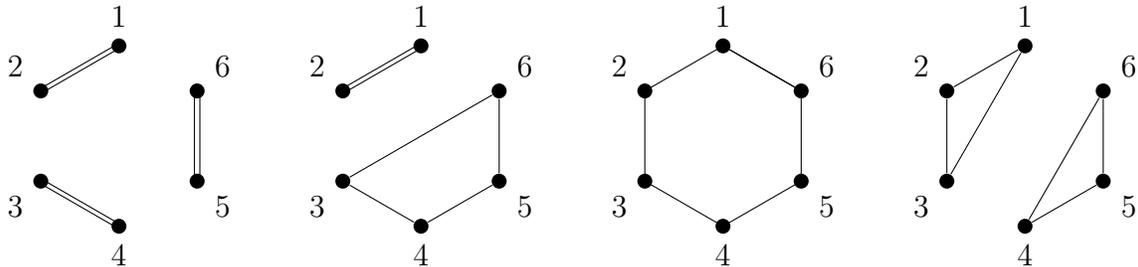

Next, we have the twenty-prime six-point function, which is of particular interest to us since we will concern ourselves with the two-loop six-point conformal integrals. For this case, the polynomial in $Y_i$ which describes this function should contain up to four different kinds of structures
\begin{equation}
	Y_{12}^2 \, Y_{34}^2 \, Y_{56}^2 \,\, , \hspace{0.5cm} Y_{12}^2 \, Y_{34} \, Y_{45} \, Y_{56} \, Y_{63} \,\, , \hspace{0.5cm} Y_{12} \, Y_{23} \, Y_{31} \, Y_{45} \, Y_{56} \, Y_{64} \,\, , \hspace{0.5cm} Y_{12} \, Y_{23} \, Y_{34} \, Y_{45} \, Y_{56} \, Y_{61} \,\, ,
	\label{eq:possible structures six pt function}
\end{equation}
which can be seen from the diagrams of figure \ref{fig:diagram structures six-point function},
together with all the possible permutations. Therefore, a reasonable choice to represent this six-point function is
\begin{equation}
     \langle \mathcal{O}_{20'}(x_1,Y_1) \, \dots \, \mathcal{O}_{20'}(x_6,Y_6) \rangle = \left(\frac{Y_{12} \, Y_{34} \, Y_{56}}{x_{12}^2 \, x_{34}^2 \, x_{56}^2}\right)^2 \, \mathcal{G}_6\left(u_i,U_i;\sigma_i,\tau_i\right) \,\, ,
	\label{eq:20' six-point function}
\end{equation}
which now depends on nine cross-ratios $u_1, \dots, u_6$ and $U_1, \dots, U_3$ defined as
\begin{equation}
	u_1 = \frac{x_{12}^2 x_{35}^2}{x_{13}^2 x_{25}^2} \,\, , \hspace{0.5cm} u_{i+1} = \left.u_i\right\vert_{x_j \rightarrow x_{j+1}} \,\, , \hspace{0.7cm} U_1 = \frac{x_{13}^2 x_{46}^2}{x_{14}^2 x_{36}^2} \,\, , \hspace{0.5cm} U_{i+1} = \left.U_i\right\vert_{x_j \rightarrow x_{j+1}} \, ,
	\label{eq:six-point cross-ratios}
\end{equation}
where the superscript in $x_j$ is taken modulo $6$, and where the respective R-symmetry cross-ratios $\sigma_1, \dots, \sigma_6$ and $\tau_1, \dots, \tau_3$ are given by the above expressions with the squared distances $x_{ij}^2$ replaced by the products of the null polarization vectors $Y_{ij}$.

Concerning the corrections of this object, the integrands are only known up to $2$ loops \cite{Eden:2010zz,Eden:2010ce,Bargheer:2022sfd}, but the respective integrals have not yet been computed up to this level. The simplest of these integrals has been studied in \cite{Morales:2022csr}, where they have not only been able to obtain its symbol but also infer that these integrals must be expressed in terms of elliptic functions. These functions, however, are not as simple as we would like. Fortunately, it has been observed for higher point functions that under certain kinematical limits the associated conformal integrals simplify. Therefore, although we might not be able to evaluate the integrals in full generality we can compute them in several simpler cases and try to learn something from it. Here we shall consider two distinct circumstances: a configuration in which the operators lie in a line, and another where adjacent points will be taken to become null separated from each other $x_{i,i+1}^2 \rightarrow 0$, also known as the light-cone limit.

It is also important to mention that the higher-point functions have interesting non-trivial properties, namely the Drukker-Plefka twist \cite{Drukker:2009sf} and the chiral algebra twist \cite{Beem:2013sza}, which impose extra constraints on the correlators. Something nice about the integrals that will be studied here is that they can be used to verify these properties. 


Furthermore, the results we will obtain can not only be used to provide useful information, such as CFT data, but they can also serve later as cross-checks. Indeed, if these integrals happen to be computed for general kinematics or by means of other methods, one can always restrict the results to the same kinematics as here and see if both results agree. This has already been the case for four- and five-point correlation functions, where the results from the evaluation of conformal integrals matched those obtained by means of integrability methods \cite{Fleury:2016ykk,Eden:2016xvg,Fleury:2017eph,Fleury:2020ykw}. In particular, integrability can be used to obtain the correlators as a power series over the cross-ratios, where generally one also considers the points to lie on a plane or some other simpler kinematics. Because the results these methods provide us are already full integrated functions, simply knowing the integrands of the $\ell$ loop corrections is not enough. As such, evaluating the conformal integrals for the six-point function, even if for rather particular kinematics, would already be useful to compare with results from integrability. Furthermore, given that from the integrability viewpoint it is hard to make an identification between a certain integrability contribution and a conformal integral, our results could also help us make predictions regarding this issue.




\section{Integrals} \label{sec:Integrals}
Our goal is to compute two-loop six-point conformal integrals of the form 
\begin{equation}
	I = \int \frac{d^d x_7 \, d^d x_8}{x_{78}^2 \prod_{i=1}^{6} \prod_{j=7}^{8} (x_{ij}^2)^{a_{ij}}} \, ,
	\label{eq:type of integrals 2 loops}
\end{equation} 
where $a_{ij}$ are integers that satisfy $\sum_{i} a_{ij} = d-1$. Below we explain a possible way of evaluating these integrals which was the one used here to compute the integrals for the line configuration.

\subsection{Methodology} \label{subsec:Methodology}
The first step of the computation is to use conformal symmetry to simplify the integrand. In particular, we send one of the external points to infinity. Although the choice of the point is completely arbitrary it is good practice to choose one of the points that appears more times in the denominator or one that appears in the numerator. Additionally, it is also useful to send one point to lie at the origin and another to a fixed position, usually chosen to be $(1,0,\dots)$.

In case the remaining integrand has no position-dependent numerator, the next step is to perform the Schwinger parametrization (see \cite{itzykson2012quantum} for example), which makes use of the following identity
\begin{equation}
	\frac{1}{A^n} = \frac{1}{\Gamma(n)} \int_{0}^{\infty} d\alpha \, \alpha^{n-1} e^{-\alpha A} \, .
	\label{eq:identity for Schwinger parametrization}
\end{equation}
Upon using this equality, the spacetime integrals transform into Gaussian integrals, which can easily be done, and we obtain integrals over the Schwinger parameters \cite{Panzer:2014caa,Panzer:2015ida}
\begin{equation}
	I(a_{ij}) = \Gamma(\omega) \, \left(\prod_{i} \int_{0}^{\infty} \frac{d\alpha_i \, \alpha_i^{a_i - 1}}{\Gamma(a_i)} \right) \, \frac{\Psi^{\omega - \frac{d}{2}}}{\Phi^{\frac{d}{2}}} \, \delta(1 - \alpha_j) \, ,
	\label{eq:Schwinger parameters representation of loop integral}
\end{equation}
where $\Psi$ and $\Phi$ are graph polynomials resulting from the spacetime integrations, the index $i$ runs over the propagators appearing in the integral (\ref{eq:type of integrals 2 loops}), with $a_i$ being their powers, and where the superficial degree of divergence $\omega$ for any number of loops $\ell$ is given by
\begin{equation}
	\omega = \sum_{i} a_i - \frac{d}{2} \cdot \ell \, .
	\label{eq:formula omega}
\end{equation}
The Dirac delta at the end fixes some arbitrary Schwinger parameter $\alpha_j$ to one.
 
Finally, the integrals over the parameters $\alpha_i$ can be evaluated by means of the computer package \texttt{HyperInt} \cite{Panzer:2015ida}, provided that we consider some simple geometry or kinematical limits which simplify the integrands. At the end, we use the function \texttt{fibrationBasis}, belonging to the same package, that expresses the result in a way that is suitable for expansion around $0$ with respect to the provided arguments and the specified order.

When it is the case that even after using conformal symmetry we still have a position-dependent numerator in the integrand the procedure is slightly different. This has to do with the fact that the identity (\ref{eq:identity for Schwinger parametrization}) has problems for negative $n$, in which case the Gamma function has poles. Nevertheless, the procedure does not differ significantly from what was just explained, requiring us to differentiate with respect to some additional Schwinger parameters and evaluate them at zero. We give a more detailed explanation of this in an example below but we refer the reader to \cite{Golz:2015rea} for a complete description.

\subsection{Two-loop half-BPS six-point integrals} \label{subsec:20' integrals}
The two-loop correction of single-trace half-BPS six-point functions involves conformal integrals of three different types
\begin{align}
	& \mathcal{B}_{123,456} \equiv \int \frac{d^4 x_7 \, d^4 x_8}{(x_{17}^2 x_{27}^2 x_{37}^2) \, x_{78}^2 \, (x_{48}^2 x_{58}^2 x_{68}^2)} \, , \label{eq:B type integral} \\
	& \mathcal{P}_{1,23,456} \equiv \int \frac{x_{28}^2 \, d^4 x_7 \, d^4 x_8}{(x _{17}^2 x_{27}^2 x_{37}^2) \, x_{78}^2 \, (x_{18}^2 x_{48}^2 x_{58}^2 x_{68}^2)} \, , \label{eq:P type integral} \\
	& \mathcal{T}_{14,23,56} \equiv \int \frac{x_{28}^2 \, x_{57}^2 \, d^4 x_7 \, d^4 x_8}{(x _{17}^2 x_{27}^2 x_{37}^2 x_{47}^2) \, x_{78}^2 \, (x_{18}^2 x_{48}^2 x_{58}^2 x_{68}^2)} \, . \label{eq:T type integral}
\end{align}
Following the same convention as \cite{Bargheer:2022sfd}\footnote{The conformal integrals presented here are the same as the ones from \cite{Bargheer:2022sfd} with the difference being that we are using a different notation and we have ignored the prefactors.} we will refer to the $\mathcal{B}$ integrals as double-box integrals, $\mathcal{P}$ as penta-box integrals and $\mathcal{T}$ as double-penta integrals. Due to the symmetries of the integrands, each of these conformal integrals are invariant under certain permutations of positions. For example, switching $1 \leftrightarrow 2$ in the $\mathcal{B}$ integral above amounts to the same integral. Consequently, the number of independent integrals for each type is: 10 $\mathcal{B}$'s, 60 $\mathcal{P}$'s and 180 $\mathcal{T}$'s. Luckily, we do not have to compute all of these integrals, which would be quite cumbersome, but only those with which we can obtain all others from cross-ratios transformations.

Ideally, we would like to be able to compute these integrals in the most general configuration of the six-point function. Unfortunately, we have not yet been able to evaluate integrals of the sort expressed in (\ref{eq:type of integrals 2 loops}) under general kinematics. Nonetheless, we can still study them in specific cases for which the integrands simplify. In principle, this allow us to compute the integrals and obtain closed form results in terms of well defined functions.

\subsection{Line geometry} \label{subsec:Line geometry}
One of the possibilities is to consider a configuration in which all points lie in the same line. We thus start by using conformal symmetry to send three points to $0$, $1$ and $\infty$, that we respectively choose to be $x_1$, $x_2$ and $x_6$, and then impose the remaining points to lie on a line by fixing $x_{ij}^2 = (z_i - z_j)(\bar{z}_i - \bar{z}_j)$ with $z_i = \bar{z}_i$. In this configuration, the cross-ratios $u_i$, $U_i$ of the six-point function (\ref{eq:six-point cross-ratios}) are related to the $z_i$ cross-ratios by
\begin{equation}
	\begin{aligned}
		u_1 = \frac{z_2^2 \, (1-z_5)^2}{(z_2 - z_5)^2} \, ,& \,\, u_2 = \frac{(z_2 - 1)^2}{(z_2 - z_4)^2} \, , \,\, u_3 = \frac{(1-z_4)^2 \, z_5^2}{z_4^2 \, (1-z^5)^2} \, , \,\, u_4 = \frac{(z_4 - z_5)^2}{(z_2 - z_5)^2} \, , \,\, u_5 = \frac{1}{z_5^2} \, , \\
		& U_1 = \frac{1}{z_4^2} \, , \,\, U_2 = \frac{(z_2 - z_4)^2 \, z_5^2}{z_4^2 \, (z_2 - z_5)^2} \, , \,\, U_3 = \frac{(1-z_5)^2}{(z_2 - z_5)^2} \, .
		\label{eq:relation u,U and zs cross-ratios collinear geometry}
	\end{aligned}
\end{equation}
Relatedly, the choice of cross-ratios with which we work can be crucial for the evaluation of the integral. To see this let us consider the 
one-loop correction to the four point function of scalars in four dimensions
\begin{equation}
	\int \frac{d^4 x_0}{x_{01}^2 \, x_{02}^2 \, x_{03}^2 \, x_{04}^2} \,\, .
	\label{eq:first order correction to four point function}
\end{equation}
Following the procedure mentioned above we start by sending $x_4 \rightarrow \infty$, we then introduce the Schwinger parameters and integrate over $x_0$. After doing this we obtain the integral
\begin{equation}
	\int \frac{d\alpha_1 d\alpha_2}{(1 + \alpha_1 + \alpha_2)(\alpha_1 + v \, \alpha_2 + u \, \alpha_1 \,\alpha_2)^2} \, ,
	\label{eq:4pts 1 loop correction in u and v}
\end{equation}
where the third Schwinger parameter has been fixed to $1$ by the Dirac delta. It is not hard to do one of the two above integrals. If we choose to do the one in $\alpha_1$ we find that this double integral gets reduced to
\begin{equation}
	\int d \alpha_2 \, \frac{1 + \left\{1 + u - v\left[1 - \log(v \alpha_2) + \log\big((1+\alpha_2)(1+ u \, \alpha_2)\big) \right]\right\} \alpha_2 + u \, \alpha_2^2}{v \, \alpha_2 \, \left[1 + \alpha_2 (1 - v + u (1 + \alpha_2))\right]^2} \, .
	\label{eq:4pts 1 loop correction in u and v after alpha1 integration}
\end{equation}
The issue with the above expression is that some of the factors in the denominator are not linear in the integration variable. We want to avoid  these kinds of terms as the method used by \texttt{HyperInt} requires the integrands to factor linearly. Generally, we can overcome this problem simply by using different variables. In this particular case if we use the cross-ratios $z$ and $\bar{z}$ instead of $u$ and $v$ the integral over Schwinger parameters is given by
\begin{equation}
	\int \frac{d\alpha_1 d\alpha_2}{(1 + \alpha_1 + \alpha_2)((1-z)(1-\bar{z})\alpha_2 + \alpha_1 (1 + z \bar{z} \alpha_2))^2} \, ,
	\label{eq:4pts 1 loop correction in z and zb}
\end{equation}
and the one we obtain after performing one of the integrations is
\begin{equation}
	\int \!\! d \alpha_2 \, \tfrac{1 \, + \, \alpha_2 \, \big\{z + \bar{z} + (1-z)\, (1-\bar{z}) \big[\log\left[(1-z) \, (1-\bar{z}) \, \alpha_2\right] - \log\left[(1+\alpha_2) \, (1 + z \bar{z} \alpha_2)\right]\big]  + z \, \bar{z} \, \alpha_2 \big\}    }{(1-z) \, (1-\bar{z}) \, \alpha_2 \, (1 + z \alpha_2)^2 \, (1 + \bar{z} \alpha_2)^2} \, .
	\label{eq:4pts 1 loop correction in z and zb after alpha1 integration}
\end{equation}
Now, we no longer have terms in $\alpha_2$ as the one found above, thus making the second integral more tractable. 

For higher-point functions the choice of variables such as not to have these non-linear terms is in general not straightforward. For the six-point integrals considered here (\ref{eq:B type integral},\ref{eq:P type integral},\ref{eq:T type integral}) there might exist some possible choice that allows us to only have linearized terms in the integrands, which in turn makes the integrals easier to evaluate. As far as we know such a choice was not yet found. Therefore, we opted to work with the $z,\bar{z}$ cross-ratios and further restrict to the line  configuration $z = \bar{z}$ in order to avoid the non-linear terms and be able to compute the integrals.

As mentioned previously, some of the integrals will still have numerator terms even after using conformal symmetry to fix some of the points. We would now like to illustrate what the procedure for these cases is when evaluating them on a line. For this, we will consider one of the double-penta integrals, namely
\begin{equation}
	\mathcal{T}_{16,23,45} = \int \frac{x_{28}^2 \, x_{47}^2 \, d^4 x_7 \, d^4 x_8}{x_{17}^2 x_{27}^2 x_{37}^2 x_{78}^2 x_{18}^2 x_{48}^2 x_{58}^2} \, ,
	\label{eq:T_16,23,45 integral}
\end{equation}
where we have already used conformal symmetry to send $x_6 \rightarrow \infty$. Following the recipe we gave before, we then use Schwinger parametrization for every term in the integrand and integrate over the Gaussian space-time integrals. Next, we restrict the points to the plane, send $x_1 \rightarrow 0$ and $x_3 \rightarrow 1$ and further restrict to the line kinematics $z_i = \bar{z}_i$. The outcome is an integral over Schwinger parameters of the form (\ref{eq:Schwinger parameters representation of loop integral}), with the graph polynomials $\Psi$ and $\Phi$ as given in appendix \ref{sec:Graph Polynomials}. Their expressions depend on nine Schwinger parameters, given that we have nine squared distances terms, but should only depend on seven since it is the number of terms on the denominator. This has to do with the aforementioned fact that the Schwinger parametrization as in (\ref{eq:identity for Schwinger parametrization}) is problematic for the numerator terms. In these cases what we must do according to \cite{Golz:2015rea} is to use the more general form of the Schwinger trick (see \cite{itzykson2012quantum} for example). This will involve taking two derivatives for each of the two extra Schwinger parameters $\alpha_8$ and $\alpha_9$ evaluated at zero. Since one of the two derivatives for each Schwinger parameter cancels the respective integration, the integral will then be given by 
\begin{equation}
		\mathcal{T}_{16,23,45} \propto \prod_{i=1}^{7} \int_{0}^{\infty} d\alpha_i  \, \delta(1 - \alpha_j) \, \frac{\partial}{\partial \alpha_8} \left[\frac{\partial}{\partial \alpha_9} \left[\frac{\Psi}{\Phi^2}\right]_{\alpha_9 = 0}\right]_{\alpha_8 = 0}   \, ,
\end{equation}
where we already used that $d = 4$ and $\omega = 1$ for this case. A small comment is in order here. In particular, although we can take the two derivatives and then evaluate the integrals, we found that in some cases the resulting integrands after the derivatives are complex enough that the integrals take a very long time to be evaluated with \texttt{HyperInt}. What one must do is to take one derivative, evaluate a few of the integrals over Schwinger parameters, say $\alpha_1$, $\alpha_3$ and $\alpha_4$ for example\footnote{The order of integration on the $\alpha$'s is also important to prevent the evaluation of the integrals from taking too long. To find the best order one uses the functions \texttt{irreducibles}, \texttt{cgReduction} and \texttt{suggestIntegrationOrder} described in \cite{Panzer:2015ida}.}, take the other derivative and only then compute the remaining integrals. This is really a trial and error game that has to be played for each integral to see what works best. At the end, we use $\texttt{fibrationBasis}$ to write the result in a way that is suitable for being expanded in the limit of small arguments.

Finally, we would just like to note that because on the line configuration we will be working with general cross-ratios $z_i$, we simply have to evaluate a conformal integral of each type. All others follow from appropriately transforming the cross-ratios according to the respective positions permutations.

\subsection{Light-cone limit and asymptotic expansions}\label{subsec:Light-cone limit}
Another choice that can be taken in order to simplify the conformal integrals is to consider physically relevant kinematical limits. In particular, limits in which the leading order terms of these integrals are relatively easier to compute. One of these cases is the light-cone limit, where we take $x_{i,i+1}^2 \rightarrow 0$. When in this kinematics the conformal integrals can be separated in distinct regions according to whether the integrations variables are small or not and use certain identities that make the integrals easier. This procedure is called asymptotic expansion method (see \cite{Smirnov:2002pj,Eden:2012rr,Goncalves:2016vir} for example). 

Let us then consider we are taking the null limit $x_{12}^2 \rightarrow 0$. The idea behind the asymptotic expansions is that we may start by separating each integration into two regions: one where the integration variable is small and thus comparable to the small parameter $x_{12}^2$, and another where this variable is big such that dropping the squared distance is a good approximation. Given that we are dealing with two-loop integrals, we will have four regions:
\begin{equation}
	\text{I.} \,\, x_7 \ll 1 , x_8 \ll 1 \, ;  \,\,\,\,\, \text{II.} \,\, x_7 \ll 1 , x_8 \gg 1 \, ; \,\,\,\,\, \text{III.} \,\, x_7 \gg 1 , x_8 \ll 1 \, ;  \,\,\,\,\, \text{IV.} \,\, x_7 \gg 1 , x_8 \gg 1 \,\, .
	\label{eq:regions of integration for light-cone limit}
\end{equation}
In general, when we take the light-cone limit between any two points, an $\ell$-loop integral will give rise to $2^\ell$ regions. After separating the integral in these four regions, it is possible to simplify the integrand by making use of the identity
\begin{equation}
	\frac{1}{(x_{ij}^2)^b} = \frac{1}{(x_{1j}^2)^b} \sum_{n=0}^{\infty} \binom{-b}{n} \, \frac{(x_{1i}^2 - 2 x_{1 i} \cdot x_{1 j})^n}{(x_{1j}^2)^n} \,\, ,
	\label{eq:identity for expansion}
\end{equation} 
where $x_i$ is assumed to be small and $x_j$ to be large. Naturally these expansions have to be done in a suitable manner for each region.

For the sake of exemplification let us then consider one of the simpler six-point conformal integrals, namely (\ref{eq:B type integral}). Following the aforementioned prescription we can write
\begin{equation}
	\begin{aligned}
		&\hspace{4.5cm}\mathcal{B}_{123,456} = \mathcal{B}_1+ \mathcal{B}_2 + \mathcal{B}_3 + \mathcal{B}_4 \, ,\\
		& \mathcal{B}_1 = \sum_{n,m,k} \int \frac{d^d x_7 d^d x_8 \, (2x_{13}\cdot x_{17} - x_{17}^2)^{n} (2x_{14}\cdot x_{18} - x_{18}^2)^{m} (2x_{15}\cdot x_{18} - x_{18}^2)^{k}}{(x_{13}^2)^{1+n} \, (x_{14}^2)^{1+m} \, (x_{15}^2)^{1+k} \, x_{17}^2 \, x_{27}^2 \, x_{78}^2} \, , \\
		& \hspace{2cm} \mathcal{B}_2 = \sum_{n,m} \int \frac{d^d x_7 d^d x_8 \, (2x_{13}\cdot x_{17} - x_{17}^2)^{n} (2x_{17}\cdot x_{18} - x_{17}^2)^{m}}{(x_{13}^2)^{1+n} \, (x_{18}^2)^{1+m} \, x_{17}^2 \, x_{27}^2 \, x_{48}^2 \, x_{58}^2} \, , \\
		&\hspace{3.5cm} \mathcal{B}_4 = \sum_{n} \int \frac{d^d x_7 d^d x_8 \, (2 x_{12} \cdot x_{17} - x_{12}^2)^{n}}{(x_{17}^2)^{2+n} \, x_{37}^2 \, x_{48}^2 \, x_{58}^2 \, x_{78}^2} \, ,
	\end{aligned}
	\label{eq:B_123,456 integral regions}
\end{equation}
where for simplicity we have used conformal symmetry to send $x_6 \rightarrow \infty$. The reason why we have not given the explicit expression for $\mathcal{B}_3$ is that it is scaleless \cite{Smirnov:2002pj,Abreu:2022mfk} and thus evaluates to zero. Moreover, we point out to the dimensional regularization $d=4-\epsilon$ performed in the integrals measures. This is a key step of the asymptotic expansion method which allows to regulate the possible divergences of each region integral. Importantly, the final result should not depend on $\epsilon$ since the conformal integrals had no dependence on it to start with.

Hopefully the above expressions convey the central idea of this method, which is that we obtain several integrals that are easier to evaluate than the original one. If needed, however, we could further take the null limits $x_{34}^2 \rightarrow 0$ and $x_{56}^2 \rightarrow 0$ and follow the same path to obtain even simpler integrals. For the integrals considered in this work we found no such thing was necessary as one null limit sufficed to evaluate them.

To compute the integrals in (\ref{eq:B_123,456 integral regions}) we can start by dropping the terms $x_{12}^2$, $x_{17}^2$ and $x_{18}^2$ inside the parenthesis terms in the numerators since they are subleading. Afterwards, we can obtain the final expressions for these integrals by making use of the following result \cite{Bercini:2024unp}
\begin{equation}
	\int \frac{d^d x_7 \, d^d x_8 \, (x_{15} \cdot x_{57})^{k_1} \, (x_{25} \cdot x_{58})^{k_2}}{x_{57}^2 x_{58}^2 x_{67}^2 x_{68}^2 x_{78}^2} = c_{k_1,k_2} \, \frac{(x_{15} \cdot x_{56})^{k_1} (x_{25} \cdot x_{56})^{k_2}}{(x_{56}^2)^{d-5}} \, ,
	\label{eq:identity for integral}
\end{equation}
with
\begin{equation}
	c_{k_1,k_2} = \binom{k_1 + k_2}{k_1} \sum_{j=0}^{k_2} (-1)^j \binom{k_2}{j} \frac{1}{1 + j + k_1} \[6 \zeta_3 + S_{1,2}(k_1 + j) - S_{2,1}(k_1 + j)\] \, ,
	\label{eq:definition of c_k1,k2}
\end{equation}
where $S_{i,j}$ are harmonic sums defined recursively as
\begin{equation}
	S_{a,\textbf{b}}(N) = \sum_{n=1}^{N} \frac{(\text{sign}(a))^{n}}{n^{|a|}} \, S_{\textbf{b}} (n) \, , \,\,\,\,\, \text{with} \,\,\, S_{\emptyset}(N) = 1 \,\, , \,\,\,\, a,b_i \in \mathbb{Z} \setminus \{0\} \, .
	\label{eq:harmonic sums}
\end{equation}
Even though our integrals might have more terms in the numerator than what appears in the equality (\ref{eq:identity for integral}), this does not prevent us from using it, as explained in appendix \ref{sec:Asymptotic expanded integrals}. 

In general, when we take kinematical limits it is no longer true that we can obtain all of the conformal integrals simply by computing one integral of each kind and appropriately transforming the cross-ratios. This stems from the fact that in these cases some of these integrals become independent of each other. Therefore, what must be done is that we have to compute those integrals with which we can generate all others in these limits. In our case, it is possible to infer that the conformal integrals we need to assess are
\begin{equation}
	\mathcal{B}_{123,456} \,\, , \hspace{0.5cm} \mathcal{B}_{135,246} \,\, ,
	\label{eq:light-cone limit B independent integrals} 
\end{equation}
for the double-box integrals,
\begin{equation}
	\mathcal{P}_{2,13,456} \,\, , \,\,\,\, \mathcal{P}_{3,12,456} \,\, , \,\,\,\, \mathcal{P}_{3,14,256} \,\, , \,\,\,\, \mathcal{P}_{3,15,246} \,\, , \,\,\,\, 
	\label{eq:light-cone limit P independent integrals}
\end{equation}
for the penta-box integrals, and 
\begin{equation}
	\mathcal{T}_{46,15,23} \,\, , \,\,\,\, \mathcal{T}_{46,15,32} \,\, , \,\,\,\, \mathcal{T}_{56,14,23} \,\, , \,\,\,\, \mathcal{T}_{56,14,32} \,\, , \,\,\,\, \mathcal{T}_{56,12,34} \,\, , \,\,\,\, \mathcal{T}_{26,15,34} \,\, , \,\,\,\, \mathcal{T}_{24,16,35} \,\, , \,\,\,\,
	\label{eq:light-cone limit T independent integrals}
\end{equation}
for the double-penta integrals \footnote{Naturally, we are free to permute the indices and obtain another set of independent integrals, although we gain nothing new.}. As it so happens, some of these integrals are easier to evaluate than others, namely $\mathcal{T}_{56,14,23}$, $\mathcal{T}_{56,14,32}$ and $\mathcal{T}_{56,12,34}$. Here we decided to compute only one of these three integrals, in particular $\mathcal{T}_{56,14,23}$, but the procedure for the other two is exactly the same. The more difficult ones, however, will be left to be computed in some future work.

Now that we have given some insights about the kinematics and respective procedures to deal with the integrals at hand, we will proceed to present the obtained results and respective comments.

\section{Results and analysis} \label{sec:Results and analysis}
\subsection{Line configuration}
We will start by analyzing the results obtained from evaluating the integrals in the line configuration. We have chosen to compute all ten double-box integrals, given that the number is fairly reasonable. However, when it comes to the penta-box integrals and the double-penta integrals, we have only evaluated the integrals $\mathcal{P}_{1,23,456}$ and $\mathcal{T}_{16,23,45}$, respectively, as it would be quite time-consuming to compute all others. Nonetheless, one can still easily obtain the remaining ones by seeing how the cross-ratios transform under indices permutations, as mentioned in section \ref{subsec:20' integrals}. For instance, under the exchange of points $2 \leftrightarrow 6$, the cross-ratios would transform as
\begin{equation}
	z_2 \rightarrow 1-z_2 \,\, , \,\,\,\, z_4 \rightarrow \frac{z_4 \, (z_2 - 1)}{z_2 - z_4} \,\, , \,\,\,\, z_5 \rightarrow \frac{z_5 \, (z_2 - 1)}{z_2 - z_5} \,\, .
	\label{eq:transformation of cross-ratios under permutation of indices}
\end{equation}
Afterwards, we simply have to expand the final expression accordingly to the kinematical limit we want to analyze.

All the integrals were computed in four different cases, such that the following kinematical regimes could be taken afterwards if one wanted, namely: $z_2, z_4, z_5 \rightarrow 0$; $z_2, z_5 \rightarrow 0$, $z_4 \rightarrow 1$; $z_2, z_4 \rightarrow 0$ and $z_5 \rightarrow \infty$;  $z_2 \rightarrow 0$, $z_4 \rightarrow 1$ and $z_5 \rightarrow \infty$ (with the remaining points fixed by conformal symmetry as mentioned before). The results can be found in an auxiliary \texttt{Mathematica} file. We point out that throughout this file we have introduced two new variables defined as
\begin{equation}
	y = 1 - z_4 \,\, , \hspace{0.5cm} w = \frac{1}{z_5} \,\, ,
	\label{eq:y and w definition}
\end{equation}
which when taken to zero allow for the kinematical limits mentioned above.

Importantly, we found that all of the results were given in terms of rational functions of the cross-ratios $z_i$, zeta functions
\begin{equation}
	\zeta (s) = \sum_{n=1}^{\infty} \frac{1}{n^s} \,\, ,
	\label{eq:zeta function definition}
\end{equation}
with $s \in \mathbb{C}$, and hyperlogarithm functions defined as
\begin{equation}
	\text{Hlog}(z,[\sigma_1,\dots,\sigma_r]) = \int_{0}^{z} \frac{dz'}{z' - \sigma} \, \text{Hlog}(z',[\sigma_2,\dots,\sigma_r]) \, , \hspace{0.5cm} \text{with} \,\,\, \text{Hlog}(z)=1 \, ,
	\label{eq:Hlog def}
\end{equation}
where $\sigma_i$,$z \in \mathbb{C}$.

Interestingly enough, we observed that in all of the results the simplest type of integrals $\mathcal{B}$ factorized as a product of a common cross-ratios prefactor and a linear combination of hyperlogarithms and zeta functions. For instance, in the case in which the limit $z_i \rightarrow 0$ ($i=2,4,5$) could be taken later, we found one of these integrals to be given by
\begin{equation}
	\mathcal{B}_{123,456} = \frac{2}{z_2 (z_2 - 1) (z_4 - z_5)} \Big(\text{Hlog} (z_2, [z_4,0,z_5]) -\text{Hlog}(z_2, [z_4,1,z_5]) + \dots \Big) \,\, .
	\label{eq:B_123,456 integral first terms}
\end{equation} 
However, in the case of the $\mathcal{P}$ and $\mathcal{T}$ integrals this property is no longer observed due to the non-existence of a common cross-ratios prefactor.

Furthermore, we noted that in all the obtained results the integrals only contained hyperlogarithms with up to three parameters in the arguments list (see (\ref{eq:Hlog def})). This implies that the results can be expressed in terms of polylogarithms, defined as
\begin{equation}
	\text{Li}_{s}(z) = \sum_{k=1}^{\infty} \frac{z^k}{k^s} \,\, , \hspace{0.5cm} \,\,\, \text{with} \,\,\,\,  \text{Li}_1 (z) = - \text{log} (1-z) \,\, ,
	\label{eq:polylogarithm def} 
\end{equation}
for $s \in \mathbb{N}$ and $z \in \mathbb{C}$. This contrasts to what would happen if the list of parameters had four or more parameters, in which case the hyperlogarithms are related with the so-called multiple polylogarithms
\begin{equation}
	\text{Li}_{s_1,\dots,s_r}(z_1,\dots,z_r) = \sum_{0<k_1<\dots<k_r}^{\infty} \frac{z_1^{k_1} \dots z_r^{k_r}}{k_1^{k_1} \dots k_r^{s_r}} \,\, , \hspace{0.5cm} \,\,\, s_1, \dots, s_r \in \mathbb{N} \,\, .
\end{equation}

From looking at expression (\ref{eq:B_123,456 integral first terms}) , in particular its prefactor, one could think that it diverges in a power-law fashion when the limit $z_2 \rightarrow 0$ is taken. If we expand the expression for small cross-ratios $z_i \rightarrow 0$ in the order of increasing index, we find that the first few orders are 
\begin{equation}
	\begin{aligned}
		 \mathcal{B}_{123,456} &\underset{z_i \rightarrow 0}{\sim} \frac{1}{z_5^2}  \left[ 2 \log(z_5^2) - 4 \log(z_2) \log(z_5) - 2 \log(z_4)^2 +  4 \log(z_2) \log(z_4) \right. \\
		& \left.   + \, 6 \log(z_5) - 2 \log(z_4) - 4 \log(z_2) + 4 \right] + \frac{2}{z_5} + \dots \, ,
	\end{aligned} 
	\label{eq:B_123,456 expansion in small z_i}
\end{equation}
where the dots denote subleading terms in the expansion. From this, it appears we have power-law divergences due to negative powers in $z_5$. However, if we consider the prefactors that these integrals must have when used for the loop corrections (see \cite{Bargheer:2022sfd} for example) these disappear and only logarithmic divergences remain. This was also verified in the results for the other aforementioned limits and types of integrals. Nevertheless, note that even these logarithmic divergences might disappear when we take the linear combinations of these integrals to give the full two-loop corrections. This, however, was not analyzed here.

\subsection{Light-cone limit}
We now discuss the results obtained from evaluating the six-point conformal integrals in the light-cone limit. In particular, we will only consider the null limit $x_{56}^2 \rightarrow 0$ and not others since as previously explained taking one null limit is sufficient in our case. Moreover, as mentioned in section \ref{subsec:Light-cone limit} three of the integrals from (\ref{eq:light-cone limit B independent integrals},\ref{eq:light-cone limit P independent integrals},\ref{eq:light-cone limit T independent integrals}) are easier to evaluate. Here we chose to analyze the double-penta integral $\mathcal{T}_{56,14,23}$, which is given by
\begin{equation}
	\mathcal{T}_{56,14,23} = \int \frac{x_{18}^2 \, x_{27}^2 \, d^d x_7 \, d^d x_8}{x _{17}^2 x_{28}^2 x_{38}^2 x_{47}^2 x_{57}^2 x_{58}^2 x_{67}^2 x_{68}^2 x_{78}^2} \, \, .
	\label{eq:T_56,14,23 integral}
\end{equation}
In order to evaluate it in the light-cone limit we follow the integration by regions that we explained above. However, for this integral we will only need to concern with the first region of (\ref{eq:regions of integration for light-cone limit}), which will be denoted by $\mathcal{T}_I$, as we will only be interested in the leading order results. Indeed, from a simple analysis of the leading order behavior in $x_{56}^2$, it is possible to infer that $\mathcal{T}_I$ diverges as $(x_{56}^2)^{-1-2\epsilon}$ in the light-cone limit, while others are subleading.

After using the identity (\ref{eq:identity for expansion}) accordingly to the assumptions of the first region, we end up with
\begin{equation}
	\begin{aligned}
		&\mathcal{T}_I = \sum_{n_1,n_2,n_3,n_4,n_5} \frac{1}{(x_{15}^2)^{n_1 + n_4} \, (x_{25}^2)^{n_2 + n_5} \, (x_{35}^2)^{n_3}} \, \times \\
		&\int \frac{d^d x_7 \, d^d x_8 \, (-2 x_{15} \cdot x_{57})^{n_1} (-2 x_{25} \cdot x_{58})^{n_2} (-2 x_{35} \cdot x_{58})^{n_3} (2 x_{15} \cdot x_{58})^{n_4} (2 x_{25} \cdot x_{57})^{n_5}}{x_{57}^2 x_{58}^2 x_{67}^2 x_{68}^2 x_{78}^2}
	\end{aligned}
	\label{eq:first region light-cone limit integral}
\end{equation}
where the sums in $n_4$ and $n_5$ run only over $\{0,1\}$\footnote{The reason why these sums run only over $\{0,1\}$ is because they come from the expansion of the numerators. Therefore, the associated binomial coefficients will be $\binom{1}{n}$, which only make sense for $0 \le n \le 1$.} and where for simplicity we have sent $x_4 \rightarrow \infty$. Moreover, we have dropped the terms $x_{57}^2$ and $x_{58}^2$ inside the parentheses in the numerator terms as they are subleading relatively to the ones we have kept.

To evaluate this integral we use (\ref{eq:identity for integral}) while following the reasoning of appendix \ref{sec:Asymptotic expanded integrals} and get to the final result
\begin{equation}
	\mathcal{T}_I = \!\!\!\!\!\!\! \sum_{n_1,n_2,n_3,n_4,n_5} \!\!\!\!\!\! \frac{(-1)^{n_1 + n_2 + n_3} \, c_{n_1 + n_5, n_2 + n_3 + n_4 }  (2 x_{15} \cdot x_{56})^{n_1 + n_4}  (2 x_{25} \cdot x_{56})^{n_2 + n_5} (2 x_{35} \cdot x_{56})^{n_3}}{(x_{15}^2)^{n_1 + n_4} \, (x_{25}^2)^{n_2 + n_5} \, (x_{35}^2)^{n_3} \, (x_{56}^2)^{5-d}} \,  .
	\label{eq:T integral region 1 final result}
\end{equation}
We may take the limit $\epsilon \rightarrow 0$ such that $(5 - d) \rightarrow 1$ and explicitly see that this integral indeed diverges as $(x_{56}^2)^{-1}$ in the light-cone limit. It could also be beneficial to express this result in terms of the six-point cross-ratios, which could render its analysis simpler. This will not be done here since it would make the expression longer and thus not possible to fit it in a single line anymore. Nevertheless, the procedure would consist on using the identity
\begin{equation}
	x_{i5} \cdot x_{56} = -\frac{1}{2} \left(x_{i5}^2 + x_{56}^2 - x_{i6}^2\right) \, ,
	\label{eq:scalar product to squared distances}
\end{equation}
where we can drop $x_{56}^2$ in the light-cone limit, and finally use the definition of the cross-ratios (\ref{eq:six-point cross-ratios}).
 
\section{Discussion} \label{sec:Discussion}
Correlation functions of $\mathcal{N}=4$ SYM, especially of half-BPS operators, have been a subject of interest for some decades as they provide us information about the CFT itself and also about its dual theory in AdS. The simpler non-trivial of these are the four-point functions, which have been studied both in the weak \cite{Eden:2000mv,Eden:2011we,Chicherin:2015edu,Georgoudis:2017meq} and strong \cite{DHoker:1999kzh,Arutyunov:2000py,Goncalves:2014ffa,Rastelli:2016nze,Alday:2017xua,Binder:2019jwn} regimes of the coupling and whose loop corrections are known up to a relatively high order. When it comes to correlators with even more points, however, not so much is known. This stems from the contrasting difficulty between knowing the integrands and actually being able to evaluate the conformal integrals which give the $\ell$-loop corrections. In the case of five-point functions, the first few orders of the corrections for some of these functions have already been obtained under certain kinematical limits \cite{Bercini:2024pya}. On the other hand, the integrands of the six-point functions are known up to $2$-loops \cite{Bargheer:2022sfd} and the respective integrals have been analyzed in \cite{Morales:2022csr}, even though they were not able to explicitly evaluate them.

Here, we obtained a closed form expression for some of the two loop six-point integrals on a line configuration, which truly amounts to knowing all of them since they are related by cross-ratios transformations. For this we have adopted the Schwinger parameterization and used the \texttt{HyperInt} package \cite{Panzer:2014caa,Panzer:2015ida}, which allows to compute Feynman integrals in the Schwinger parameters representation. We have found in all the obtained results that the final expressions were given by rational functions of the cross-ratios $z_i$, hyperlogarithm functions and the zeta functions. Moreover, we took the kinematical limits mentioned in section \ref{sec:Results and analysis} for some results and observed that the results possess divergences. As was commented there, these divergences might cancel out when we sum the integrals in order to obtain the two-loop corrections, but this was not verified here. 

In addition, we also evaluated one of the double-penta integrals in the kinematics where two adjacent points become null separated. Although we only considered one of the three integrals which we claimed were easier to compute in this kinematical limit the remaining ones can be obtained by following the exact same path. In order to evaluate this integral we implemented the asymptotic expansion method. Despite the level of complexity surely rising, it would be nice to compute the remaining independent integrals (\ref{eq:light-cone limit B independent integrals},\ref{eq:light-cone limit P independent integrals},\ref{eq:light-cone limit T independent integrals}) as they are crucial if we want to obtain the actual two-loop corrections and extract physical information.

As was mentioned several times throughout this paper, we chose to consider the line configuration and the light-cone limit in order to simplify the integrands and be able to evaluate the integrals. This stems from the high complexity of the same integrals for general configurations, which so far has prevented us from computing them. It would be nice if we could somehow work around this problem and obtain the results for a general disposition of the six-point function. Naturally, this ought to be quite challenging so we could start off by taking one point out of the line and do the same analysis. We would then continue to take more points out of the line and gradually move on to more general but harder configurations and assess these integrals. 

Finally, we could try to tackle these issues from an integrability-based approach, as was done for four- and five-point functions \cite{Fleury:2016ykk,Eden:2016xvg,Fleury:2017eph,Fleury:2020ykw}, and compare the outcomes with our results. 

\section*{Acknowledgements} 
We would like to thank Vasco Gonçalves for proposing this project and for many useful discussions and also Bruno Fernandes for several insights provided throughout this work and for reviewing the manuscript. We are also grateful to João Vilas Boas for reviewing the manuscript. Centro de Física do Porto is partially funded by Fundação para a Ciência e a Tecnologia (FCT) under the grant UID04650-FCUP. The author is supported by Simons Foundation grant \#488637
(Simons collaboration on the non-perturbative bootstrap).

\appendix
\section{Graph Polynomials $\Psi$ and $\Phi$} \label{sec:Graph Polynomials}
Here we give the explicit expressions for the graph polynomials $\Phi$ and $\Psi$ that are obtained after writing the double-penta integral $\mathcal{T}_{16,23,45}$ in the Schwinger parameters representation (\ref{eq:Schwinger parameters representation of loop integral}). In particular, these are given by
\begin{equation}
	\begin{aligned}
		\Psi &= \alpha _1 \alpha _2+\alpha _1 \alpha _5+\alpha _1 \alpha _6+\alpha _1 \alpha _7+\alpha _1 \alpha _8+\alpha _2 \alpha _3 +\alpha _2 \alpha _4 +\alpha _2 \alpha _7 +\alpha _2 \alpha _9+\alpha _3 \alpha _5\\
		&+\alpha _3 \alpha _6+\alpha _3 \alpha _7+\alpha _3 \alpha _8+\alpha _4 \alpha _5+\alpha _4 \alpha _6+\alpha _4 \alpha _7 +\alpha _4 \alpha _8+\alpha _5 \alpha _7+\alpha _5 \alpha _9+\alpha _6 \alpha _7\\
		&+\alpha _6 \alpha _9+\alpha _7 \alpha _8+\alpha _7 \alpha _9+\alpha _8 \alpha _9 \, ,
	\end{aligned}
	\label{eq:Psi graph polynomial for T_16,23,45}
\end{equation}
\begin{align}
	\Phi &= \alpha _1 \alpha _2 \alpha _3 z_2^2+\alpha _2 \alpha _3 \alpha _4 z_2^2+\alpha _1 \alpha _3 \alpha _5 z_2^2+\alpha _3 \alpha _4 \alpha _5 z_2^2+\alpha _1 \alpha _3 \alpha _6 z_2^2+\alpha _3 \alpha _4 \alpha _6 z_2^2 \label{eq:Phi graph polynomial for T_16,23,45} \\
	&+\alpha _1 \alpha _3 \alpha _7 z_2^2+\alpha _2 \alpha _3 \alpha _7 z_2^2+\alpha _3 \alpha _4 \alpha _7 z_2^2+\alpha _3 \alpha _5 \alpha _7 z_2^2+\alpha _3 \alpha _6 \alpha _7 z_2^2+\alpha _1 \alpha _2 \alpha _8 z_2^2 \nonumber\\
	&+\alpha _1 \alpha _3 \alpha _8 z_2^2+\alpha _2 \alpha _3 \alpha _8 z_2^2+\alpha _2 \alpha _4 \alpha _8 z_2^2+\alpha _3 \alpha _4 \alpha _8 z_2^2+\alpha _1 \alpha _5 \alpha _8 z_2^2+\alpha _3 \alpha _5 \alpha _8 z_2^2 \nonumber\\
	&+\alpha _4 \alpha _5 \alpha _8 z_2^2+\alpha _1 \alpha _6 \alpha _8 z_2^2+\alpha _3 \alpha _6 \alpha _8 z_2^2+\alpha _4 \alpha _6 \alpha _8 z_2^2+\alpha _1 \alpha _7 \alpha _8 z_2^2+\alpha _2 \alpha _7 \alpha _8 z_2^2 \nonumber\\
	& +\alpha _4 \alpha _7 \alpha _8 z_2^2+\alpha _5 \alpha _7 \alpha _8 z_2^2+\alpha _6 \alpha _7 \alpha _8 z_2^2+\alpha _2 \alpha _3 \alpha _9 z_2^2+\alpha _3 \alpha _5 \alpha _9 z_2^2+\alpha _3 \alpha _6 \alpha _9 z_2^2 \nonumber\\
	& +\alpha _3 \alpha _7 \alpha _9 z_2^2+\alpha _2 \alpha _8 \alpha _9 z_2^2+\alpha _3 \alpha _8 \alpha _9 z_2^2+\alpha _5 \alpha _8 \alpha _9 z_2^2 +\alpha _6 \alpha _8 \alpha _9 z_2^2+\alpha _7 \alpha _8 \alpha _9 z_2^2 \nonumber\\
	& -2 \alpha _2 \alpha _3 \alpha _4 z_2  -2 \alpha _3 \alpha _4 \alpha _5 z_2-2 \alpha _3 \alpha _4 \alpha _6 z_2 -2 \alpha _3 \alpha _4 \alpha _7 z_2 -2 z_4 \alpha _3 \alpha _5 \alpha _7 z_2 \nonumber\\
	&-2 z_5 \alpha _3 \alpha _6 \alpha _7 z_2-2 \alpha _3 \alpha _4 \alpha _8 z_2-2 z_4 \alpha _1 \alpha _5 \alpha _8 z_2-2 z_4 \alpha _3 \alpha _5 \alpha _8 z_2-2 z_4 \alpha _4 \alpha _5 \alpha _8 z_2 \nonumber\\
	& -2 z_5 \alpha _1 \alpha _6 \alpha _8 z_2 -2 z_5 \alpha _3 \alpha _6 \alpha _8 z_2-2 z_5 \alpha _4 \alpha _6 \alpha _8 z_2-2 \alpha _4 \alpha _7 \alpha _8 z_2-2 z_4 \alpha _5 \alpha _7 \alpha _8 z_2 \nonumber\\
	&-2 z_5 \alpha _6 \alpha _7 \alpha _8 z_2-2 z_4 \alpha _2 \alpha _3 \alpha _9 z_2-2 z_4 \alpha _3 \alpha _5 \alpha _9 z_2-2 z_4 \alpha _3 \alpha _6 \alpha _9 z_2-2 z_4 \alpha _3 \alpha _7 \alpha _9 z_2 \nonumber\\
	&-2 z_4 \alpha _3 \alpha _8 \alpha _9 z_2-2 z_4 \alpha _5 \alpha _8 \alpha _9 z_2-2 z_5 \alpha _6 \alpha _8 \alpha _9 z_2-2 z_4 \alpha _7 \alpha _8 \alpha _9 z_2+\alpha _1 \alpha _2 \alpha _4 \nonumber\\
	&+\alpha _2 \alpha _3 \alpha _4+z_4^2 \alpha _1 \alpha _2 \alpha _5+z_4^2 \alpha _2 \alpha _3 \alpha _5+\alpha _1 \alpha _4 \alpha _5+z_4^2 \alpha _2 \alpha _4 \alpha _5+\alpha _3 \alpha _4 \alpha _5+z_5^2 \alpha _1 \alpha _2 \alpha _6 \nonumber\\
	& +z_5^2 \alpha _2 \alpha _3 \alpha _6+\alpha _1 \alpha _4 \alpha _6+z_5^2 \alpha _2 \alpha _4 \alpha _6+\alpha _3 \alpha _4 \alpha _6 + z_4^2 \alpha _1 \alpha _5 \alpha _6+z_5^2 \alpha _1 \alpha _5 \alpha _6 \nonumber\\
	& -2 z_4 z_5 \alpha _1 \alpha _5 \alpha _6+z_4^2 \alpha _3 \alpha _5 \alpha _6+z_5^2 \alpha _3 \alpha _5 \alpha _6-2 z_4 z_5 \alpha _3 \alpha _5 \alpha _6+z_4^2 \alpha _4 \alpha _5 \alpha _6+z_5^2 \alpha _4 \alpha _5 \alpha _6 \nonumber\\
	& -2 z_4 z_5 \alpha _4 \alpha _5 \alpha _6+\alpha _1 \alpha _4 \alpha _7+\alpha _2 \alpha _4 \alpha _7+\alpha _3 \alpha _4 \alpha _7 +z_4^2 \alpha _1 \alpha _5 \alpha _7 +z_4^2 \alpha _2 \alpha _5 \alpha _7 \nonumber\\
	&+z_4^2 \alpha _3 \alpha _5 \alpha _7+z_4^2 \alpha _4 \alpha _5 \alpha _7-2 z_4 \alpha _4 \alpha _5 \alpha _7+\alpha _4 \alpha _5 \alpha _7+z_5^2 \alpha _1 \alpha _6 \alpha _7 +z_5^2 \alpha _2 \alpha _6 \alpha _7 \nonumber
\end{align}
\begin{align}
	& +z_5^2 \alpha _3 \alpha _6 \alpha _7+z_5^2 \alpha _4 \alpha _6 \alpha _7-2 z_5 \alpha _4 \alpha _6 \alpha _7+\alpha _4 \alpha _6 \alpha _7+z_4^2 \alpha _5 \alpha _6 \alpha _7+z_5^2 \alpha _5 \alpha _6 \alpha _7 \nonumber \\
	&-2 z_4 z_5 \alpha _5 \alpha _6 \alpha _7+\alpha _1 \alpha _4 \alpha _8+\alpha _3 \alpha _4 \alpha _8+z_4^2 \alpha _1 \alpha _5 \alpha _8+z_4^2 \alpha _3 \alpha _5 \alpha _8+z_4^2 \alpha _4 \alpha _5 \alpha _8 \nonumber \\
	&+z_5^2 \alpha _1 \alpha _6 \alpha _8+z_5^2 \alpha _3 \alpha _6 \alpha _8+z_5^2 \alpha _4 \alpha _6 \alpha _8+\alpha _4 \alpha _7 \alpha _8+z_4^2 \alpha _5 \alpha _7 \alpha _8+z_5^2 \alpha _6 \alpha _7 \alpha _8 \nonumber\\
	& +z_4^2 \alpha _1 \alpha _2 \alpha _9 +z_4^2 \alpha _2 \alpha _3 \alpha _9 +z_4^2 \alpha _2 \alpha _4 \alpha _9-2 z_4 \alpha _2 \alpha _4 \alpha _9+\alpha _2 \alpha _4 \alpha _9+z_4^2 \alpha _1 \alpha _5 \alpha _9 \nonumber\\
	&+z_4^2 \alpha _2 \alpha _5 \alpha _9+z_4^2 \alpha _3 \alpha _5 \alpha _9+z_4^2 \alpha _4 \alpha _5 \alpha _9-2 z_4 \alpha _4 \alpha _5 \alpha _9+\alpha _4 \alpha _5 \alpha _9+z_4^2 \alpha _1 \alpha _6 \alpha _9 \nonumber\\
	&+z_5^2 \alpha _2 \alpha _6 \alpha _9+z_4^2 \alpha _3 \alpha _6 \alpha _9+z_4^2 \alpha _4 \alpha _6 \alpha _9-2 z_4 \alpha _4 \alpha _6 \alpha _9+\alpha _4 \alpha _6 \alpha _9+z_4^2 \alpha _5 \alpha _6 \alpha _9 \nonumber\\
	&+z_5^2 \alpha _5 \alpha _6 \alpha _9-2 z_4 z_5 \alpha _5 \alpha _6 \alpha _9+z_4^2 \alpha _1 \alpha _7 \alpha _9+z_4^2 \alpha _2 \alpha _7 \alpha _9+z_4^2 \alpha _3 \alpha _7 \alpha _9+z_4^2 \alpha _4 \alpha _7 \alpha _9 \nonumber\\
	&-2 z_4 \alpha _4 \alpha _7 \alpha _9+\alpha _4 \alpha _7 \alpha _9+z_4^2 \alpha _6 \alpha _7 \alpha _9+z_5^2 \alpha _6 \alpha _7 \alpha _9-2 z_4 z_5 \alpha _6 \alpha _7 \alpha _9+z_4^2 \alpha _1 \alpha _8 \alpha _9 \nonumber\\
	&+z_4^2 \alpha _3 \alpha _8 \alpha _9+z_4^2 \alpha _4 \alpha _8 \alpha _9-2 z_4 \alpha _4 \alpha _8 \alpha _9+\alpha _4 \alpha _8 \alpha _9+z_4^2 \alpha _5 \alpha _8 \alpha _9+z_5^2 \alpha _6 \alpha _8 \alpha _9+z_4^2 \alpha _7 \alpha _8 \alpha _9 \, . \nonumber
\end{align}

\section{Asymptotic expanded integrals}\label{sec:Asymptotic expanded integrals}
When evaluating the integral (\ref{eq:first region light-cone limit integral}) in the light-cone limit $x_{56}^2 \rightarrow 0$ we claimed that we could make use of the equality (\ref{eq:identity for integral}), even though our integral has more terms than what appeared in this identity. Here we will briefly explain the basic idea of why such result can be used for our case, as it merely involves making a generalization.

Let us then consider just the terms which involve $x_{57}$ and forget about the ones with $x_{58}$, where the argument is the same. Suppose then that we have more than one term of the kind appearing in our integral. For simplicity let us consider we have two of these terms
\begin{equation}
	(k_1 \cdot x_{57})^{a_1} \, (k_2 \cdot x_{57})^{a_2} \, ,
	\label{eq:product of terms appearing in light-cone integral}
\end{equation}
where $k_1$, $k_2$ and $a_1$, $a_2$ are arbitrary vectors and integers, respectively. Now, we introduce some parameter $q$ which allows us to keep track of the coefficients, just as it happens when working with generating functions. Moreover, we shall consider the sum over the exponents just as we have in our case but with a small caveat, we shall introduce by force specific coefficients which will allow us to use the following equality
\begin{equation}
	\sum_{a_1, a_2 = 0}^{\infty} \binom{a_1 + a_2}{a_2} (k_1 \cdot x_{57})^{a_1} (k_2 \cdot x_{57})^{a_2} q^{a_2} = \sum_{n=0}^{\infty} \big((k_1 + q \, k_2) \cdot x_{57}\big)^{n} \, . 
	\label{eq:identity to merge powers of scalar products}
\end{equation}

Next, we can consider the integral of this equation with the remaining factors, such as the denominator and the terms with $x_{58}$, which will be omitted for simplicity. However, since we have reduced the two terms with $x_{57}$ to just one we can use the identity (\ref{eq:identity for integral}) on the right-hand side. Schematically, this would give
\begin{equation}
	\sum_{n=0}^{\infty} c_{n,b} \, \big((k_1 + q \, k_2) \cdot x_{56}\big)^n \, \dots \, ,
	\label{eq:schematic result after integration}
\end{equation}
where the dots refer to the other terms we are omitting and $b$ denotes the power of the term with $x_{58}$, which was also integrated. In order to compare with the initial expression, i.e. the left-hand side of (\ref{eq:product of terms appearing in light-cone integral}), we use the binomial expansion
\begin{equation}
	\sum_{n=0}^{\infty} \sum_{m=0}^{n} c_{n,b} \, \binom{n}{m} (k_1 \cdot x_{56})^{n-m} \, (k_2 \cdot x_{56})^m \, q^m \, \dots \, ,
	\label{eq:binomial expansion of integral result}
\end{equation}
and we reorganize the sum by making the transformations $n \rightarrow a_1 + a_2$ and $m \rightarrow a_2$, obtaining therefore
\begin{equation}
	\sum_{a_1, a_2 = 0}^{\infty} c_{a_1 + a_2, b} \, \binom{a_1 + a_2}{a_2} (k_1 \cdot x_{56})^{a_1} (k_2 \cdot x_{56})^{a_2} q^{a_2} \, \dots \, .
	\label{eq:result after integrating}
\end{equation}
This result implies that when we have more than one term of the kind $(k \cdot x_{56})^a$, the outcome of the integration is similar to that of (\ref{eq:identity for integral}) but where in the labels of the coefficients $c_{k_1,k_2}$ (\ref{eq:definition of c_k1,k2}) appear the sums of the powers of these terms. Finally, we would just like to point out that this reasoning generalizes to any number of these terms as well as for the ones with $x_{58}$.

\newpage
\bibliographystyle{utphys}
\bibliography{TwoLoopSixPointIntegralsBibliography}

\providecommand{\href}[2]{#2}\begingroup\raggedright\begin{thebibliography}{10}

\bibitem{Eden:2010zz}
B.~Eden, G.~P. Korchemsky, and E.~Sokatchev, ``{From correlation functions to
  scattering amplitudes},''
  \href{http://dx.doi.org/10.1007/JHEP12(2011)002}{{\em JHEP} {\bfseries 12}
  (2011) 002}, \href{http://arxiv.org/abs/1007.3246}{{\ttfamily arXiv:1007.3246
  [hep-th]}}.

\bibitem{Alday:2010zy}
L.~F. Alday, B.~Eden, G.~P. Korchemsky, J.~Maldacena, and E.~Sokatchev, ``{From
  correlation functions to Wilson loops},''
  \href{http://dx.doi.org/10.1007/JHEP09(2011)123}{{\em JHEP} {\bfseries 09}
  (2011) 123}, \href{http://arxiv.org/abs/1007.3243}{{\ttfamily arXiv:1007.3243
  [hep-th]}}.

\bibitem{Eden:2010ce}
B.~Eden, G.~P. Korchemsky, and E.~Sokatchev, ``{More on the duality
  correlators/amplitudes},''
  \href{http://dx.doi.org/10.1016/j.physletb.2012.02.014}{{\em Phys. Lett. B}
  {\bfseries 709} (2012) 247--253},
  \href{http://arxiv.org/abs/1009.2488}{{\ttfamily arXiv:1009.2488 [hep-th]}}.

\bibitem{Eden:2011yp}
B.~Eden, P.~Heslop, G.~P. Korchemsky, and E.~Sokatchev, ``{The
  super-correlator/super-amplitude duality: Part I},''
  \href{http://dx.doi.org/10.1016/j.nuclphysb.2012.12.015}{{\em Nucl. Phys. B}
  {\bfseries 869} (2013) 329--377},
  \href{http://arxiv.org/abs/1103.3714}{{\ttfamily arXiv:1103.3714 [hep-th]}}.

\bibitem{Maldacena:1997re}
J.~M. Maldacena, ``{The Large N limit of superconformal field theories and
  supergravity},'' \href{http://dx.doi.org/10.4310/ATMP.1998.v2.n2.a1}{{\em
  Adv. Theor. Math. Phys.} {\bfseries 2} (1998) 231--252},
  \href{http://arxiv.org/abs/hep-th/9711200}{{\ttfamily arXiv:hep-th/9711200}}.

\bibitem{Eden:2000mv}
B.~Eden, C.~Schubert, and E.~Sokatchev, ``{Three loop four point correlator in
  N=4 SYM},'' \href{http://dx.doi.org/10.1016/S0370-2693(00)00515-3}{{\em Phys.
  Lett. B} {\bfseries 482} (2000) 309--314},
  \href{http://arxiv.org/abs/hep-th/0003096}{{\ttfamily arXiv:hep-th/0003096}}.

\bibitem{Eden:2011we}
B.~Eden, P.~Heslop, G.~P. Korchemsky, and E.~Sokatchev, ``{Hidden symmetry of
  four-point correlation functions and amplitudes in N=4 SYM},''
  \href{http://dx.doi.org/10.1016/j.nuclphysb.2012.04.007}{{\em Nucl. Phys. B}
  {\bfseries 862} (2012) 193--231},
  \href{http://arxiv.org/abs/1108.3557}{{\ttfamily arXiv:1108.3557 [hep-th]}}.

\bibitem{Chicherin:2015edu}
D.~Chicherin, J.~Drummond, P.~Heslop, and E.~Sokatchev, ``{All three-loop
  four-point correlators of half-BPS operators in planar $ \mathcal{N} $ = 4
  SYM},'' \href{http://dx.doi.org/10.1007/JHEP08(2016)053}{{\em JHEP}
  {\bfseries 08} (2016) 053}, \href{http://arxiv.org/abs/1512.02926}{{\ttfamily
  arXiv:1512.02926 [hep-th]}}.

\bibitem{Georgoudis:2017meq}
A.~Georgoudis, V.~Goncalves, and R.~Pereira, ``{Konishi OPE coefficient at the
  five loop order},'' \href{http://dx.doi.org/10.1007/JHEP11(2018)184}{{\em
  JHEP} {\bfseries 11} (2018) 184},
  \href{http://arxiv.org/abs/1710.06419}{{\ttfamily arXiv:1710.06419
  [hep-th]}}.

\bibitem{DHoker:1999kzh}
E.~D'Hoker, D.~Z. Freedman, S.~D. Mathur, A.~Matusis, and L.~Rastelli,
  ``{Graviton exchange and complete four point functions in the AdS / CFT
  correspondence},''
  \href{http://dx.doi.org/10.1016/S0550-3213(99)00525-8}{{\em Nucl. Phys. B}
  {\bfseries 562} (1999) 353--394},
  \href{http://arxiv.org/abs/hep-th/9903196}{{\ttfamily arXiv:hep-th/9903196}}.

\bibitem{Arutyunov:2000py}
G.~Arutyunov and S.~Frolov, ``{Four point functions of lowest weight CPOs in
  N=4 SYM(4) in supergravity approximation},''
  \href{http://dx.doi.org/10.1103/PhysRevD.62.064016}{{\em Phys. Rev. D}
  {\bfseries 62} (2000) 064016},
  \href{http://arxiv.org/abs/hep-th/0002170}{{\ttfamily arXiv:hep-th/0002170}}.

\bibitem{Goncalves:2014ffa}
V.~Gon\c{c}alves, ``{Four point function of $\mathcal{N}=4$ stress-tensor
  multiplet at strong coupling},''
  \href{http://dx.doi.org/10.1007/JHEP04(2015)150}{{\em JHEP} {\bfseries 04}
  (2015) 150}, \href{http://arxiv.org/abs/1411.1675}{{\ttfamily arXiv:1411.1675
  [hep-th]}}.

\bibitem{Rastelli:2016nze}
L.~Rastelli and X.~Zhou, ``{Mellin amplitudes for $AdS_5\times S^5$},''
  \href{http://dx.doi.org/10.1103/PhysRevLett.118.091602}{{\em Phys. Rev.
  Lett.} {\bfseries 118} no.~9, (2017) 091602},
  \href{http://arxiv.org/abs/1608.06624}{{\ttfamily arXiv:1608.06624
  [hep-th]}}.

\bibitem{Alday:2017xua}
L.~F. Alday and A.~Bissi, ``{Loop Corrections to Supergravity on $AdS_5 \times
  S^5$},'' \href{http://dx.doi.org/10.1103/PhysRevLett.119.171601}{{\em Phys.
  Rev. Lett.} {\bfseries 119} no.~17, (2017) 171601},
  \href{http://arxiv.org/abs/1706.02388}{{\ttfamily arXiv:1706.02388
  [hep-th]}}.

\bibitem{Binder:2019jwn}
D.~J. Binder, S.~M. Chester, S.~S. Pufu, and Y.~Wang, ``{$ \mathcal{N} $ = 4
  Super-Yang-Mills correlators at strong coupling from string theory and
  localization},'' \href{http://dx.doi.org/10.1007/JHEP12(2019)119}{{\em JHEP}
  {\bfseries 12} (2019) 119}, \href{http://arxiv.org/abs/1902.06263}{{\ttfamily
  arXiv:1902.06263 [hep-th]}}.

\bibitem{Bargheer:2022sfd}
T.~Bargheer, T.~Fleury, and V.~Gon\c{c}alves, ``{Higher-point integrands in
  $\mathcal{N} = 4$ super Yang-Mills theory},''
  \href{http://dx.doi.org/10.21468/SciPostPhys.15.2.059}{{\em SciPost Phys.}
  {\bfseries 15} no.~2, (2023) 059},
  \href{http://arxiv.org/abs/2212.03773}{{\ttfamily arXiv:2212.03773
  [hep-th]}}.

\bibitem{Bercini:2024pya}
C.~Bercini, B.~Fernandes, and V.~Gon\c{c}alves, ``{Two loop five point
  integrals: light, heavy and large spin correlators},''
  \href{http://arxiv.org/abs/2401.06099}{{\ttfamily arXiv:2401.06099
  [hep-th]}}.

\bibitem{Goncalves:2019znr}
V.~Gon\c{c}alves, R.~Pereira, and X.~Zhou, ``{$20'$ Five-Point Function from
  $AdS_5\times S^5$ Supergravity},''
  \href{http://dx.doi.org/10.1007/JHEP10(2019)247}{{\em JHEP} {\bfseries 10}
  (2019) 247}, \href{http://arxiv.org/abs/1906.05305}{{\ttfamily
  arXiv:1906.05305 [hep-th]}}.

\bibitem{Goncalves:2023oyx}
V.~Gon\c{c}alves, C.~Meneghelli, R.~Pereira, J.~Vilas~Boas, and X.~Zhou,
  ``{Kaluza-Klein five-point functions from AdS$_{5}$\texttimes{}S$^{5}$
  supergravity},'' \href{http://dx.doi.org/10.1007/JHEP08(2023)067}{{\em JHEP}
  {\bfseries 08} (2023) 067}, \href{http://arxiv.org/abs/2302.01896}{{\ttfamily
  arXiv:2302.01896 [hep-th]}}.

\bibitem{Intriligator:1998ig}
K.~A. Intriligator, ``{Bonus symmetries of N=4 superYang-Mills correlation
  functions via AdS duality},''
  \href{http://dx.doi.org/10.1016/S0550-3213(99)00242-4}{{\em Nucl. Phys. B}
  {\bfseries 551} (1999) 575--600},
  \href{http://arxiv.org/abs/hep-th/9811047}{{\ttfamily arXiv:hep-th/9811047}}.

\bibitem{Caron-Huot:2010ryg}
S.~Caron-Huot, ``{Notes on the scattering amplitude / Wilson loop duality},''
  \href{http://dx.doi.org/10.1007/JHEP07(2011)058}{{\em JHEP} {\bfseries 07}
  (2011) 058}, \href{http://arxiv.org/abs/1010.1167}{{\ttfamily arXiv:1010.1167
  [hep-th]}}.

\bibitem{Eden:2000bk}
B.~Eden, A.~C. Petkou, C.~Schubert, and E.~Sokatchev, ``{Partial
  nonrenormalization of the stress tensor four point function in N=4 SYM and
  AdS / CFT},'' \href{http://dx.doi.org/10.1016/S0550-3213(01)00151-1}{{\em
  Nucl. Phys. B} {\bfseries 607} (2001) 191--212},
  \href{http://arxiv.org/abs/hep-th/0009106}{{\ttfamily arXiv:hep-th/0009106}}.

\bibitem{Nirschl:2004pa}
M.~Nirschl and H.~Osborn, ``{Superconformal Ward identities and their
  solution},'' \href{http://dx.doi.org/10.1016/j.nuclphysb.2005.01.013}{{\em
  Nucl. Phys. B} {\bfseries 711} (2005) 409--479},
  \href{http://arxiv.org/abs/hep-th/0407060}{{\ttfamily arXiv:hep-th/0407060}}.

\bibitem{Eden:2012tu}
B.~Eden, P.~Heslop, G.~P. Korchemsky, and E.~Sokatchev, ``{Constructing the
  correlation function of four stress-tensor multiplets and the four-particle
  amplitude in N=4 SYM},''
  \href{http://dx.doi.org/10.1016/j.nuclphysb.2012.04.013}{{\em Nucl. Phys. B}
  {\bfseries 862} (2012) 450--503},
  \href{http://arxiv.org/abs/1201.5329}{{\ttfamily arXiv:1201.5329 [hep-th]}}.

\bibitem{Bourjaily:2016evz}
J.~L. Bourjaily, P.~Heslop, and V.-V. Tran, ``{Amplitudes and Correlators to
  Ten Loops Using Simple, Graphical Bootstraps},''
  \href{http://dx.doi.org/10.1007/JHEP11(2016)125}{{\em JHEP} {\bfseries 11}
  (2016) 125}, \href{http://arxiv.org/abs/1609.00007}{{\ttfamily
  arXiv:1609.00007 [hep-th]}}.

\bibitem{Drummond:2013nda}
J.~Drummond, C.~Duhr, B.~Eden, P.~Heslop, J.~Pennington, and V.~A. Smirnov,
  ``{Leading singularities and off-shell conformal integrals},''
  \href{http://dx.doi.org/10.1007/JHEP08(2013)133}{{\em JHEP} {\bfseries 08}
  (2013) 133}, \href{http://arxiv.org/abs/1303.6909}{{\ttfamily arXiv:1303.6909
  [hep-th]}}.

\bibitem{Alday:2017vkk}
L.~F. Alday and S.~Caron-Huot, ``{Gravitational S-matrix from CFT dispersion
  relations},'' \href{http://dx.doi.org/10.1007/JHEP12(2018)017}{{\em JHEP}
  {\bfseries 12} (2018) 017}, \href{http://arxiv.org/abs/1711.02031}{{\ttfamily
  arXiv:1711.02031 [hep-th]}}.

\bibitem{Alday:2018pdi}
L.~F. Alday, A.~Bissi, and E.~Perlmutter, ``{Genus-One String Amplitudes from
  Conformal Field Theory},''
  \href{http://dx.doi.org/10.1007/JHEP06(2019)010}{{\em JHEP} {\bfseries 06}
  (2019) 010}, \href{http://arxiv.org/abs/1809.10670}{{\ttfamily
  arXiv:1809.10670 [hep-th]}}.

\bibitem{Caron-Huot:2018kta}
S.~Caron-Huot and A.-K. Trinh, ``{All tree-level correlators in
  AdS$_{5}$\texttimes{}S$_{5}$ supergravity: hidden ten-dimensional conformal
  symmetry},'' \href{http://dx.doi.org/10.1007/JHEP01(2019)196}{{\em JHEP}
  {\bfseries 01} (2019) 196}, \href{http://arxiv.org/abs/1809.09173}{{\ttfamily
  arXiv:1809.09173 [hep-th]}}.

\bibitem{Morales:2022csr}
R.~Morales, A.~Spiering, M.~Wilhelm, Q.~Yang, and C.~Zhang, ``{Bootstrapping
  Elliptic Feynman Integrals Using Schubert Analysis},''
  \href{http://dx.doi.org/10.1103/PhysRevLett.131.041601}{{\em Phys. Rev.
  Lett.} {\bfseries 131} no.~4, (2023) 041601},
  \href{http://arxiv.org/abs/2212.09762}{{\ttfamily arXiv:2212.09762
  [hep-th]}}.

\bibitem{Drukker:2009sf}
N.~Drukker and J.~Plefka, ``{Superprotected n-point correlation functions of
  local operators in N=4 super Yang-Mills},''
  \href{http://dx.doi.org/10.1088/1126-6708/2009/04/052}{{\em JHEP} {\bfseries
  04} (2009) 052}, \href{http://arxiv.org/abs/0901.3653}{{\ttfamily
  arXiv:0901.3653 [hep-th]}}.

\bibitem{Beem:2013sza}
C.~Beem, M.~Lemos, P.~Liendo, W.~Peelaers, L.~Rastelli, and B.~C. van Rees,
  ``{Infinite Chiral Symmetry in Four Dimensions},''
  \href{http://dx.doi.org/10.1007/s00220-014-2272-x}{{\em Commun. Math. Phys.}
  {\bfseries 336} no.~3, (2015) 1359--1433},
  \href{http://arxiv.org/abs/1312.5344}{{\ttfamily arXiv:1312.5344 [hep-th]}}.

\bibitem{Fleury:2016ykk}
T.~Fleury and S.~Komatsu, ``{Hexagonalization of Correlation Functions},''
  \href{http://dx.doi.org/10.1007/JHEP01(2017)130}{{\em JHEP} {\bfseries 01}
  (2017) 130}, \href{http://arxiv.org/abs/1611.05577}{{\ttfamily
  arXiv:1611.05577 [hep-th]}}.

\bibitem{Eden:2016xvg}
B.~Eden and A.~Sfondrini, ``{Tessellating cushions: four-point functions in
  $\mathcal{N} $ = 4 SYM},''
  \href{http://dx.doi.org/10.1007/JHEP10(2017)098}{{\em JHEP} {\bfseries 10}
  (2017) 098}, \href{http://arxiv.org/abs/1611.05436}{{\ttfamily
  arXiv:1611.05436 [hep-th]}}.

\bibitem{Fleury:2017eph}
T.~Fleury and S.~Komatsu, ``{Hexagonalization of Correlation Functions II:
  Two-Particle Contributions},''
  \href{http://dx.doi.org/10.1007/JHEP02(2018)177}{{\em JHEP} {\bfseries 02}
  (2018) 177}, \href{http://arxiv.org/abs/1711.05327}{{\ttfamily
  arXiv:1711.05327 [hep-th]}}.

\bibitem{Fleury:2020ykw}
T.~Fleury and V.~Goncalves, ``{Decagon at Two Loops},''
  \href{http://dx.doi.org/10.1007/JHEP07(2020)030}{{\em JHEP} {\bfseries 07}
  (2020) 030}, \href{http://arxiv.org/abs/2004.10867}{{\ttfamily
  arXiv:2004.10867 [hep-th]}}.

\bibitem{itzykson2012quantum}
C.~Itzykson and J.~Zuber, {\em Quantum Field Theory}.
\newblock Dover Books on Physics. Dover Publications, 2012.
\newblock \url{https://books.google.pt/books?id=CxYCMNrUnTEC}.

\bibitem{Panzer:2014caa}
E.~Panzer, ``{Algorithms for the symbolic integration of hyperlogarithms with
  applications to Feynman integrals},''
  \href{http://dx.doi.org/10.1016/j.cpc.2014.10.019}{{\em Comput. Phys.
  Commun.} {\bfseries 188} (2015) 148--166},
  \href{http://arxiv.org/abs/1403.3385}{{\ttfamily arXiv:1403.3385 [hep-th]}}.

\bibitem{Panzer:2015ida}
E.~Panzer, \href{http://dx.doi.org/10.18452/17157}{{\em {Feynman integrals and
  hyperlogarithms}}}.
\newblock PhD thesis, Humboldt U., 2015.
\newblock \href{http://arxiv.org/abs/1506.07243}{{\ttfamily arXiv:1506.07243
  [math-ph]}}.

\bibitem{Golz:2015rea}
M.~Golz, E.~Panzer, and O.~Schnetz, ``{Graphical functions in parametric
  space},'' \href{http://dx.doi.org/10.1007/s11005-016-0935-6}{{\em Lett. Math.
  Phys.} {\bfseries 107} no.~6, (2017) 1177--1192},
  \href{http://arxiv.org/abs/1509.07296}{{\ttfamily arXiv:1509.07296
  [math-ph]}}.

\bibitem{Smirnov:2002pj}
V.~A. Smirnov, ``{Applied asymptotic expansions in momenta and masses},'' {\em
  Springer Tracts Mod. Phys.} {\bfseries 177} (2002) .

\bibitem{Eden:2012rr}
B.~Eden, ``{Three-loop universal structure constants in N=4 susy Yang-Mills
  theory},'' \href{http://arxiv.org/abs/1207.3112}{{\ttfamily arXiv:1207.3112
  [hep-th]}}.

\bibitem{Goncalves:2016vir}
V.~Gon\c{c}alves, ``{Extracting OPE coefficient of Konishi at four loops},''
  \href{http://dx.doi.org/10.1007/JHEP03(2017)079}{{\em JHEP} {\bfseries 03}
  (2017) 079}, \href{http://arxiv.org/abs/1607.02195}{{\ttfamily
  arXiv:1607.02195 [hep-th]}}.

\bibitem{Abreu:2022mfk}
S.~Abreu, R.~Britto, and C.~Duhr, ``{The SAGEX review on scattering amplitudes
  Chapter 3: Mathematical structures in Feynman integrals},''
  \href{http://dx.doi.org/10.1088/1751-8121/ac87de}{{\em J. Phys. A} {\bfseries
  55} no.~44, (2022) 443004}, \href{http://arxiv.org/abs/2203.13014}{{\ttfamily
  arXiv:2203.13014 [hep-th]}}.

\bibitem{Bercini:2024unp}
C.~Bercini, B.~Fernandes, and V.~Gon\c{c}alves. {Unpublished notes}.

\end{thebibliography}\endgroup

\end{document}